\documentclass[aps,rmp,twocolumn,amsmath,amssymb,showpacs]{revtex4-1}

\usepackage[utf8]{inputenc}
\usepackage{graphicx}
\usepackage{textcomp}
\usepackage{amsmath}
\usepackage{amssymb}
\usepackage{siunitx}
\usepackage{hyperref}
\usepackage{xcolor}
\newcommand{\ket}[1]{|#1\rangle}

\begin{document}

\title{Cavity-enhanced quantum network nodes}

\author{Andreas Reiserer}
\email{andreas.reiserer@mpq.mpg.de}
\affiliation{Max-Planck-Institut f\"ur Quantenoptik, Hans-Kopfermann-Strasse 1, D-85748 Garching, and Munich Center for Quantum Science and Technology (MCQST), Ludwig-Maximilians-Universit\"at
M\"unchen - Fakult\"at f\"ur Physik, Schellingstr. 4, D-80799 M\"unchen, Germany}

\begin{abstract}
A future quantum network will consist of quantum processors that are connected by quantum channels, just like conventional computers are wired up to form the Internet. In contrast to classical devices, however, the entanglement and non-local correlations available in a quantum-controlled system facilitate novel fundamental tests of quantum theory. In addition, they enable numerous applications in distributed quantum information processing, quantum communication, and precision measurement. 

While pioneering experiments have demonstrated the entanglement of two quantum nodes separated by up to $\SI{1.3}{\kilo\meter}$, and three nodes in the same laboratory, accessing the full potential of quantum networks requires scaling of these prototypes to many more nodes and global distances. This is an outstanding challenge, posing high demands on qubit control fidelity, qubit coherence time, and coupling efficiency between stationary and flying qubits.

In this work, I will describe how optical resonators facilitate quantum network nodes that achieve the above-mentioned prerequisites in different physical systems --- trapped atoms, defect centers in wide-bandgap semiconductors, and rare-earth dopants --- by enabling high-fidelity qubit initialization and readout, efficient generation of qubit-photon and remote qubit-qubit entanglement, as well as quantum gates between stationary and flying qubits. These advances open a realistic perspective towards the implementation of global-scale quantum networks in the near future.
\end{abstract}

\maketitle

\tableofcontents

\section{Introduction}

The field of quantum technology aims at harnessing the strangeness and the power of quantum mechanics in order to implement devices that provide functionalities which are unattainable for any classical machine. In the last decades, four main fields of applications have been identified: First, quantum communication \cite{gisin_quantum_2007, ekert_ultimate_2014}, which e.g. allows for \emph{provably} secure encryption and authentication without any assumptions about the capabilities of an adversary. Second, quantum computation \cite{preskill_quantum_2018}, which can fundamentally increase the size or speed upon solving specific computational tasks. Third, quantum simulation \cite{georgescu_quantum_2014}, in which complex and inaccessible quantum systems are emulated on a device that is more accessible, with the aim to gain a better understanding and to eventually guide the development of e.g. new materials and drugs. Finally, quantum sensing \cite{degen_quantum_2017} that can improve the resolution or sensitivity of measurements.

\begin{figure}[htb!]
\centering
\includegraphics[width=1. \columnwidth]{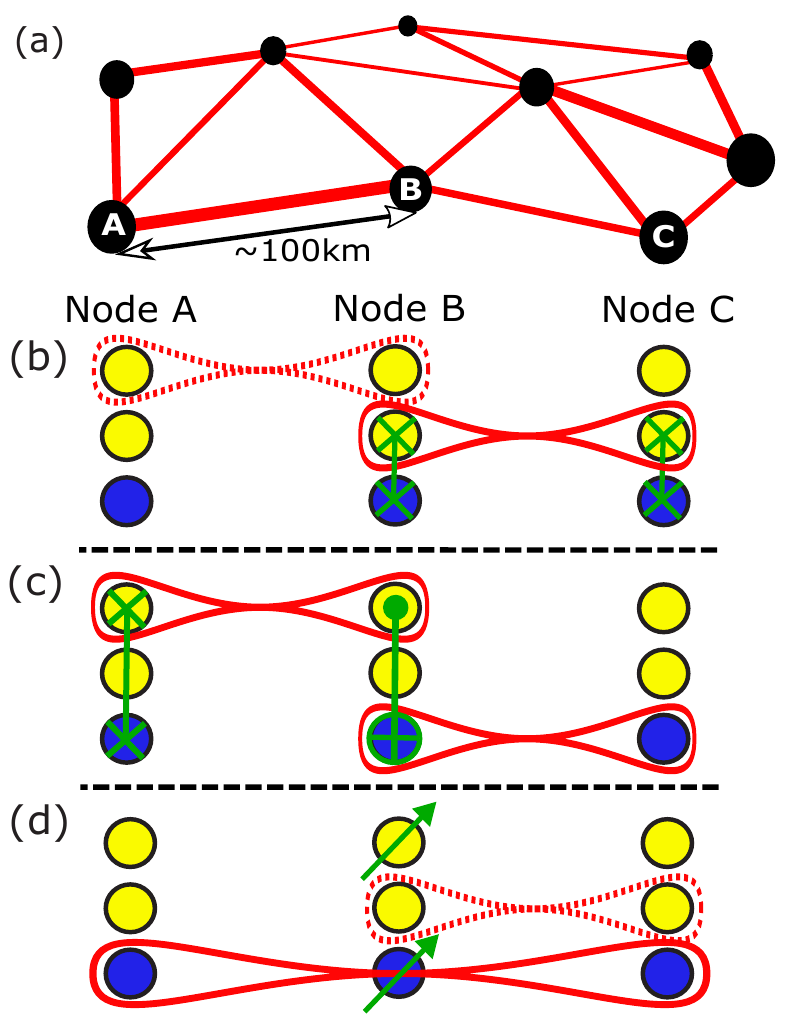}
\caption{\label{fig_repeater}
\textbf{Quantum network and quantum repeater scheme.} (a) In a quantum network, qubits at stationary nodes (black) are connected by photons that travel along optical channels (red/grey). Photon loss limits the distance of direct links to $\sim 100\,\si{\kilo\meter}$. (b) This limit can be overcome in a network architecture with memory and communication qubits. To this end, each node is equipped with several qubits (filled circles). A subset, called ‘communication’ qubits (top), couples to the optical channels for probabilistic entanglement generation (dashed symbol) between remote nodes. A herald unambiguously tells when an entanglement attempt was successful (solid symbol). In this case, local two-qubit operations are used to swap the state of the communication qubits to the ‘memory’ qubits (bottom), which are isolated from the optical channels.  (c) When memory qubits in neighboring segments have been entangled using a repeat-until-success strategy, local deterministic operations at the intermediate node can generate a maximally entangled cluster state of many qubits across the network. (d) A suited measurement of the inner qubits (arrows) can remove them from the cluster, generating an entangled state of memory qubits at the outer nodes even if their separation is too large for a direct photonic connection. Arbitrarily increasing the number or distance of entangled qubits is hampered by the loss of fidelity, which is exponential in the number of imperfect operations. To overcome this, new entanglement can be generated using the communication qubits (dashed), which may be then used to implement an error-correction layer, e.g. in the form of entanglement distillation (not shown).
}
\end{figure}

The above applications have vastly different requirements with respect to the isolation, coherence and techniques to control the used quantum system. Therefore, often specialized hardware is employed. As an example, optical photons facilitate the distribution of quantum states at the fastest possible speed -- the speed of light. Furthermore, they can be coupled into optical fibers and transmitted with negligible decoherence over many kilometers before they get absorbed.  Finally, since photons do not interact, they can be multiplexed to the same channel to achieve higher rates. The above-mentioned properties make photons ideally suited for quantum communication \cite{gisin_quantum_2007, ekert_ultimate_2014}, but are a severe disadvantage for all applications that require to keep quantum information over longer times , or to interact with other qubits or fields. This includes sensors of stationary fields \cite{degen_quantum_2017},  as well as processors and memory elements of a quantum computer \cite{preskill_quantum_2018}. Therefore, for the latter tasks other physical systems seem favorable. Most prominently, the spin of atoms in vacuum, or impurities and dopants in certain solids, offers unrivaled coherence time. Unfortunately, the isolation required for such long-term memory impedes the efficient and controlled coupling between qubits, as required for information processing.

In a hybrid system of light and matter qubits, forming a quantum network \cite{duan_colloquium:_2010, reiserer_cavity-based_2015} or "quantum internet" \cite{kimble_quantum_2008, wehner_quantum_2018}, one can achieve the above-mentioned contradicting requirements of implementing a controlled coupling between qubits while isolating them from the environment. The realization of such a quantum network may thus be an enabling technology for applications in all fields of quantum science: In quantum communication, entanglement-assisted communication can ensure unbreakable encryption \cite{ekert_ultimate_2014} and facilitate other important tasks such as authentication, position verification, secret sharing, voting, and compression \cite{buhrman_nonlocality_2010, wehner_quantum_2018}. In addition, a network of distributed quantum sensors may measure time \cite{komar_quantum_2014}, magnetic fields, gravity, or starlight \cite{gottesman_longer-baseline_2012, khabiboulline_optical_2019} with unprecedented sensitivity or resolution \cite{proctor_multiparameter_2018}. Furthermore, in quantum computing and simulation a modular architecture may improve scalability \cite{monroe_scaling_2013, awschalom_quantum_2013, kinos_roadmap_2021} by connecting smaller processing units via photons. In such remote systems, one can avoid crosstalk and correlated errors that can be difficult to correct \cite{lidar_quantum_2013}. Finally, quantum networks may allow users with finite quantum capabilities to perform computations on a remote quantum supercomputer \cite{barz_demonstration_2012, fitzsimons_private_2017}.

In addition to the known applications, novel possibilities of unforeseeable impact may emerge once global quantum networks become available. This puts the realization of a scalable quantum network at the forefront of today’s quantum science. 

In this context, scalability means that adding another entangled node or increasing the distance between the nodes will add technical complexity and require additional resources, but will not be hindered by fundamental restrictions. In current physical systems, however, there exist two fundamental restrictions that have to be overcome: Absorption that is unavoidable in any quantum channel, and errors caused by decoherence and control imperfections.

The former is the major challenge for scaling to larger distances: Consider an optical fiber link between distant nodes, operating at a telecommunication wavelength where the loss is lowest \cite{lines_search_1984}, such that the probability of transmitting a photon decreases exponentially with distance by only 0.2 dB/km. Assuming that one can realize a source of single photons with unity efficiency and the highest imaginable repetition rate, say $1\,\si{\tera\hertz}$, the success rate will drop from $7\,\si{\giga\hertz}$ after $100\,\si{\kilo\meter}$ to once every 164 years after $1000\,\si{\kilo\meter}$. Clearly, such low rates hinder quantum secure communication and the extension of quantum networks to global distances. 

But not only the success rate, also the quality of entangled states decreases exponentially when increasing the distance or the number of entangled particles. The reason is that all operations required to control qubits -- both locally and remote -- suffer from decoherence and technical imperfections, which accumulate with increasing number of qubits and operations. Thus, the realization of large-scale quantum networks will require suited protocols that counteract the accumulation of such imperfections. Such protocols are often termed "quantum error correction"\cite{devitt_quantum_2013}.

A first idea how the mentioned challenges can be overcome in order to scale quantum networks to global distances was developed in 1998 \cite{briegel_quantum_1998} in the seminal quantum repeater protocol, whose basic idea is explained in Fig. \ref{fig_repeater}. The proposed scheme involves the following key elements: First, probabilistic but heralded remote entanglement used in a repeat-until-success strategy; second, network nodes equipped with several multiplexed qubits, which can be individually controlled and coupled by deterministic operations; and third, a layer of quantum error correction, originally in the form of nested entanglement distillation \cite{bennett_purification_1996}.

The requirement of two-way signaling in the original scheme can be overcome using quantum error correction \cite{devitt_quantum_2013} instead of entanglement distillation. Then, even protocols without long-lived quantum memories can be envisioned \cite{munro_quantum_2012}, which may facilitate improved rates in certain parameter regimes, but require very high-quality operations and low optical loss \cite{muralidharan_optimal_2016}. In contrast, variants of the original scheme can be realized with experimental parameters that are accessible in the near term \cite{rozpedek_parameter_2018}. Still, achieving the required multiplexing capacity and satisfying the high demands on efficiency and fidelity of all operations is a formidable experimental challenge. In this work, I will first explain why the integration of qubits into optical resonators opens promising perspectives to this end. I will then summarize the state of the art in cavity-enhanced quantum network nodes in the most promising experimental platforms studied so far. Finally, I will give an outlook to the future prospects and challenges of these systems.

\section{Cavity-enhanced quantum network nodes}

\begin{figure}[t!] \centering
\includegraphics[width=1. \columnwidth]{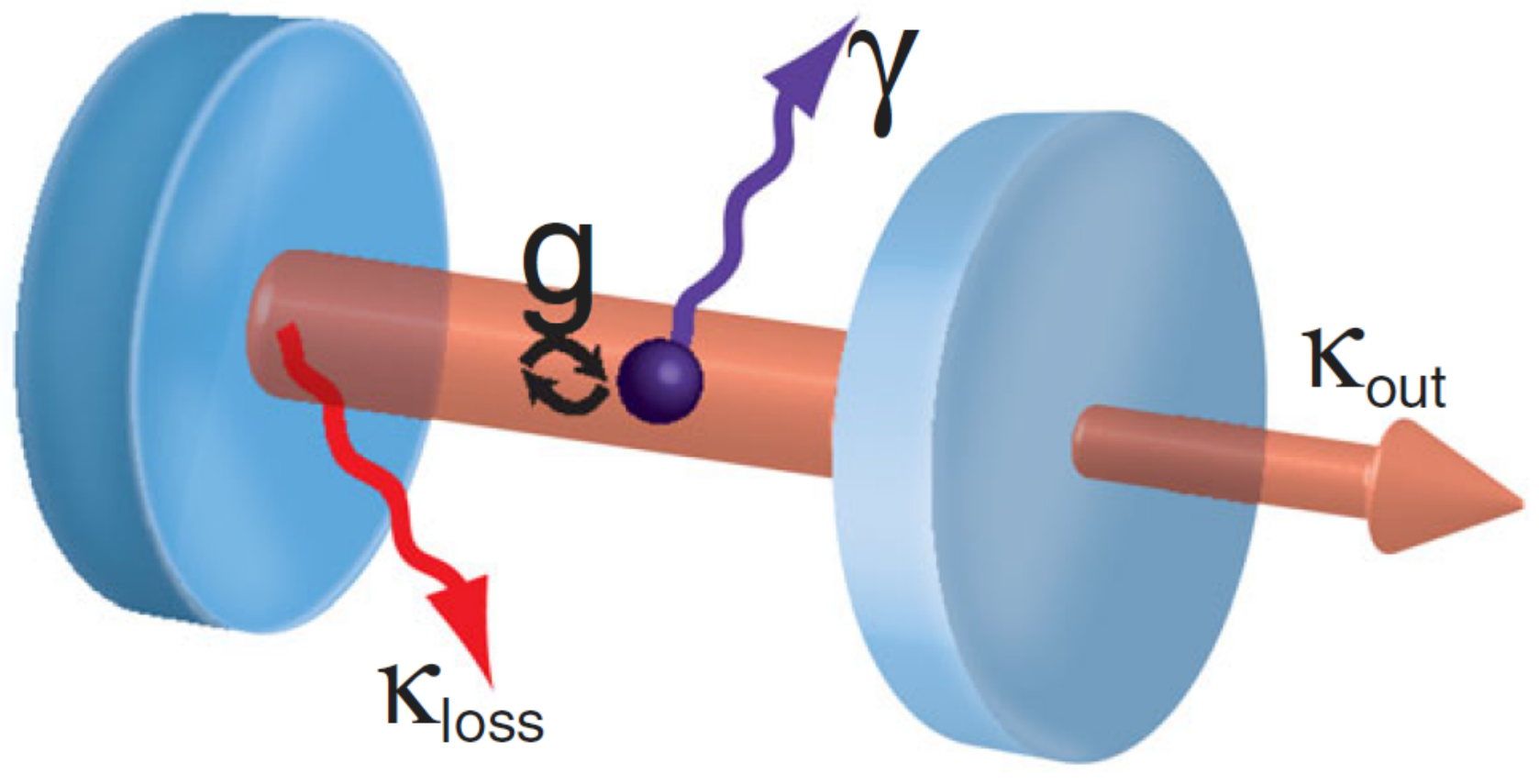}
\caption{\label{fig_EmitterInCavity}
\textbf{Single emitter in an optical resonator}. The electromagnetic field of a photon is tightly confined by two mirrors. A single emitter is located at the field maximum. The relevant rates are the emitter-cavity coupling ($g$), the decay of the emitter by spontaneous emission (at rate $2\gamma$), and the field decay of the cavity both into free-space or absorbed modes ($\kappa_\text{loss}$) and into a desired output mode ($\kappa_\text{out}$).}
\end{figure}

The experimental realization of quantum networks requires stationary nodes with qubits that can be initialized, manipulated, entangled and read individually with high fidelity. As shown in Fig. \ref{fig_repeater}, this allows for the implementation of repeater architectures \cite{briegel_quantum_1998}, in which the nodes hold two types of qubits: First, memory qubits that can store quantum states much longer than the time it takes to distribute entanglement over the network. To achieve this, the memory qubits should be decoupled from the optical channel, but exhibit a deterministic and controlled coupling mechanism to the second qubit type called communication qubits. The main purpose of the latter is in turn to provide an efficient and coherent interface to optical photons. A natural candidate for the communication qubits are single emitters, such as trapped ions \cite{duan_colloquium:_2010}, neutral atoms \cite{reiserer_cavity-based_2015}, quantum dots \cite{lodahl_interfacing_2015, gao_coherent_2015}, molecules \cite{wang_turning_2019}, and spin qubits in solid state host materials \cite{awschalom_quantum_2018, atature_material_2018, zhong_emerging_2019, wolfowicz_quantum_2021}. Often modeled as two-level systems, such emitters naturally exhibit nonlinear couplings \cite{chang_quantum_2014}. This allows for deterministic two-qubit quantum gates within the nodes, which can be a paramount advantage compared to quantum networking protocols that only use quantum memories and linear optics, which will be discussed in Sec. \ref{sec_ensembles}.

\begin{figure*}[ht!] \centering
\includegraphics[width=2. \columnwidth]{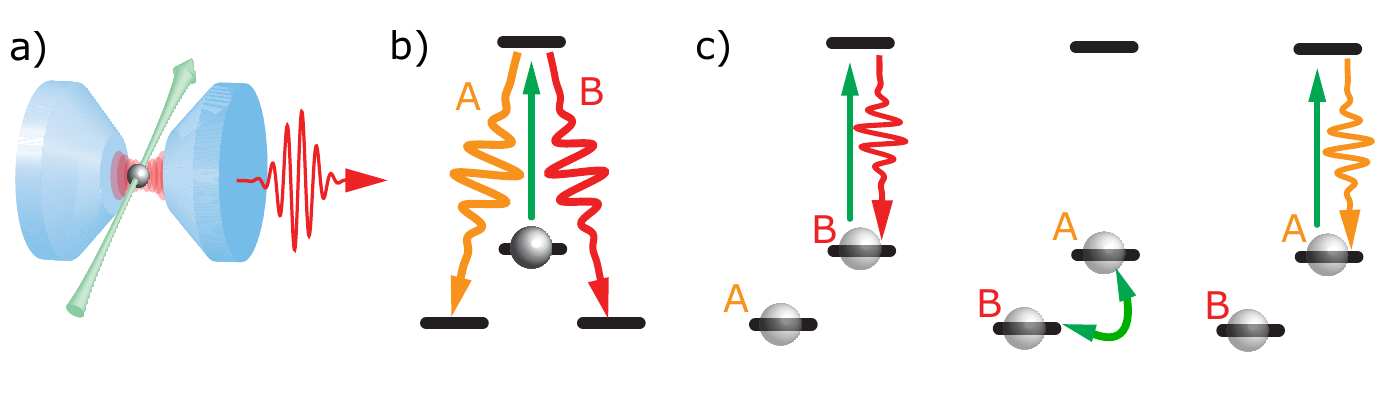}
\caption{\label{fig_Entanglement}
\textbf{Generation of spin photon entanglement}. a) A single emitter is excited with a laser pulse (arrow). It decays back to the ground state while emitting a single photon (curly arrow), which is efficiently collected by the resonator. b) Level scheme with two optical transitions. After optical excitation (straight arrow), the emitter can decay to two different levels (A and B) by emitting light at different polarization or frequency. If the transitions have equal probability and are both supported by the cavity, a maximally entangled Bell state is generated. c) Protocol with a single optical transition. The qubit is prepared in a superposition between the two ground states (A and B). In a first step (left), the emitter is excited and emits a photon only if it is in state B. After swapping the ground-state populations, a second pulse excites the spin only if it is in A. For a balanced superposition, the scheme generates entanglement between the spin state and the emission time bin of a photonic qubit.
}
\end{figure*}

The intrinsic nonlinearity of single emitters comes at the price of a moderate coupling to the photonic channels. Consider a single two-level atom that interacts with a resonant single-photon light pulse in free space \cite{leuchs_lightmatter_2013}. To achieve the best coupling, the light field would be focused to a diffraction-limited spot, of the order of $A=(\lambda/2)^2$, with $\lambda$ denoting the optical wavelength. To estimate the interaction probability, this has to be compared to the absorption cross section of the emitter, $\sigma_\text{abs}=3\lambda^2/2\pi$. Albeit $A \simeq \sigma_\text{abs}$ in this idealized situation, the photon absorption or scattering probability is in practice limited to about $20\,\%$ due to finite solid angle coverage, imperfect spatial, temporal, and polarization mode matching \cite{leuchs_lightmatter_2013}. Coupling to several levels, which is present and strong in most emitters under study, leads to a further reduction. Therefore, better confinement of the electromagnetic field of the photon in both space and time is desirable for efficient quantum network nodes. This can be achieved by tailored nanophotonic waveguides \cite{lodahl_interfacing_2015, vetsch_optical_2010}, or by embedding the emitter into an optical resonator, which is the focus of this work. 

In such a scenario, the physics of the coupled system is described by the Jaynes-Cummings Hamiltonian \cite{jaynes_comparison_1963}, as detailed in many textbooks \cite{haroche_exploring_2013}. The relevant figure of merit for the light-matter interaction and its dynamics is the cooperativity $C=g^2/(2\kappa\gamma)$ which is determined by three quantities: First, the coupling constant $g$. Second, the polarization decay rate of the emitter $\gamma=\gamma_0+\gamma_1+\gamma_d$, which stems from of its spontaneous decay on the cavity-coupled optical (at rate $2\gamma_0$) or other transitions ($2\gamma_1$), as well as its dephasing rate $\gamma_d$. Finally, the cavity field decay rate $\kappa$, which is the sum of the decay into free space $\kappa_\text{loss}$ and into a desired output mode $\kappa_\text{out}$, as shown schematically in Fig. \ref{fig_EmitterInCavity}. Note that other texts (e.g. \cite{janitz_cavity_2020}) define the energy rather than the field decay rates, which gives factors of 2 in all equations.

Often, one further distinguishes two regimes: That of "strong coupling", where $g \gg \kappa, \gamma$ and coherent reabsorption of photons is possible, and the "fast-cavity" regime with $\kappa > g \gg \gamma$ and $C \gg 1$. In both regimes, one gains access to efficient or even deterministic qubit-photon interactions \cite{borregaard_quantum_2019}, an invaluable resource for quantum networking \cite{reiserer_cavity-based_2015} and repeaters of the first generation \cite{briegel_quantum_1998}, and an indispensable prerequisite for high-rate quantum networks based on one-way quantum repeaters \cite{muralidharan_optimal_2016}. The main advantages provided by optical resonators will be described in the following sections: enhanced photon generation and absorption, improved spin-state initialization and readout, as well as spin-photon quantum gates.

\subsection{Spin-photon entanglement generation} \label{sec_SpinPhotonEntanglement}

In many protocols, the first step to entangle remote quantum network nodes is to entangle the spin of the communication qubits with single photons or photonic cluster states that are then sent along the optical channel \cite{reiserer_cavity-based_2015, borregaard_quantum_2019}. In free space, photon generation is typically realized by exciting the communication qubits with a short laser pulse, which is followed by spontaneous emission. If the emitter can decay via two transitions, the polarization of the photons can be entangled with the ground state spin level \cite{blinov_observation_2004}, as shown in Fig. \ref{fig_Entanglement}b. Alternatively, for emitters with only a single transition, entanglement with the emission time-bin is achieved using a suited sequence of ground state spin manipulations \cite{barrett_efficient_2005, bernien_heralded_2013} (see Fig. \ref{fig_Entanglement}c). In both settings, the obtained fidelity often depends on the excitation pulse duration, as the emitter can already decay during the pulse \cite{fischer_signatures_2017}, thus projecting the state or leaving the intended initial state. While this limitation is reduced with short pulses, their use is often impeded by the requirement not to drive unwanted transitions to other excited state levels.

Similar to the free-space scenario, photon generation by excitation with a short resonant laser pulse can also be implemented when the emitter is placed in a resonator (see Fig. \ref{fig_Entanglement}a). In case the emitter dephasing is not the dominant rate, which is the typical situation in quantum networking experiments, the dynamics of the decay will be strongly modified: As the density of photonic modes is changed by the resonator, one can obtain an increased \cite{purcell_spontaneous_1946} or decreased \cite{kleppner_inhibited_1981} radiative decay rate \cite{haroche_exploring_2013}:

\begin{equation}
    \gamma_c=\frac{g^2\kappa}{\kappa^2+\Delta^2}
\end{equation}

As one can see, the decay into the resonator $\gamma_c$ is suppressed when $\Delta$, the detuning between emitter and cavity mode, is increased. On resonance, one finds that the decay rate is enhanced by the Purcell effect \cite{purcell_spontaneous_1946} $\gamma_c=P\gamma_0$, where the Purcell factor $P$ is related to the cooperativity via:

\begin{equation} 
    P=2C\frac{\gamma}{\gamma_0}.
\end{equation}

In the limit of large Purcell factor, $\gamma_c \gg \gamma_0+\gamma_1$, such that the radiative decay of the emitter into the resonator is much faster than its decay into free-space modes. This has several beneficial effects in the context of quantum networks: First, when the resonator is overcoupled, i.e. the cavity field decay $\kappa$ is dominated by the coupling into a single propagating mode $\kappa_\text{out}$, one can strongly improve the photon collection probability and thus the efficiency of spin-photon entanglement generation. Second, one can enhance the photon generation rate, which is particularly relevant for emitters with slow radiative decay. But also for emitters that exhibit fast dephasing or considerable spectral diffusion, the increased decay rate can dramatically improve the coherence and spectral purity that is required for remote entanglement. A third advantage of the resonator is that it may enhance the emission into one out of several optical transitions, e.g. into a desired atomic ground state (see Sec. \ref{sec_atoms}) or crystal field level \cite{liu_spectroscopic_2005} (see Sec. \ref{sec_rare_earth_dopants}). Similarly, some emitters (see Sec. \ref{sec_defects}) exhibit an undesired co-emission of phonons whose contribution can be suppressed by the cavity-enhanced decay. Finally, the presence of a resonator enables efficient photon generation and photon absorption via off-resonant Raman transitions, which is detailed in the following.

\subsection{Stimulated Raman transitions} \label{sec_RamanTransitions}

The above-mentioned scheme of photon generation by fast resonant excitation of a two-level system is applicable to any quantum emitter, but has intrinsic limitations to the spin-photon entanglement fidelity caused by the mentioned emission during the pulse \cite{fischer_signatures_2017}. In addition, the laser pulse has to be well-separated from the single-photon pulses by spatial \cite{bochmann_fast_2008}, temporal or polarization filtering \cite{bernien_heralded_2013}. To improve the fidelity, one can use emitters with another ground-state level in a lambda configuration \cite{wilk_entanglement_2010}. Here, scattered pump light can be filtered spectrally \cite{sipahigil_integrated_2016}, and the photon emission frequency can be widely tuned \cite{mucke_generation_2013, sipahigil_integrated_2016}. In addition, re-excitation of the emitter is avoided when the ground-state level spacing is sufficient, enabling spin-photon entanglement with high fidelity (e.g. $\gtrsim 99\,\%$ at $\sim 30\,\%$ success probability) \cite{ritter_elementary_2012}, even beyond typical error correction fidelity thresholds in topological quantum computing \cite{nickerson_topological_2013, fowler_surface_2012}.

\begin{figure}[tb!]
\centering
\includegraphics[width=1. \columnwidth]{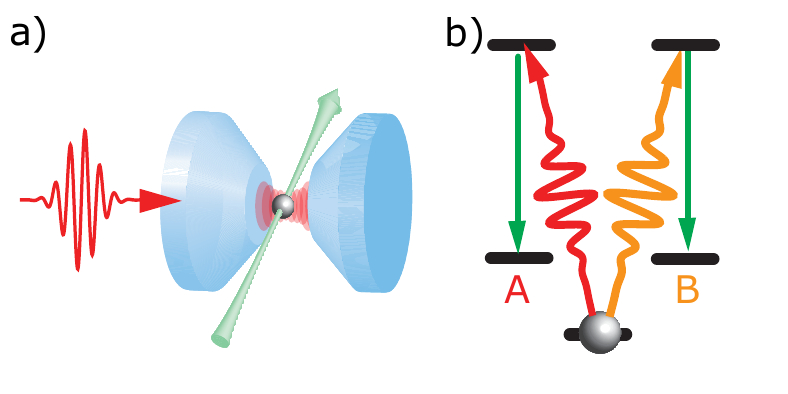}
\caption{\label{fig_Raman}
\textbf{Photon absorption using a stimulated Raman transition.} (a) An impinging photon (curly arrow) is transferred to the spin of a single emitter in a cavity using a Raman laser beam (straight arrow). (b) Level scheme. Depending on the qubit encoded in the photon, e.g. in its polarization or frequency, the emitter is transferred to a different internal state (A or B).
}
\end{figure}

When the emitter is placed in a cavity, spectral filtering is intrinsically implemented by the resonator. Even more important, photon generation is made efficient and reversible \cite{cirac_quantum_1997} when using a scheme called vacuum-stimulated Raman adiabatic passage,  pioneered with trapped atoms \cite{kuhn_controlled_1999} and later adapted to cavity-coupled solid-state emitters \cite{sun_single-photon_2018}. The scheme can be implemented both in the Purcell and in the strong-coupling regime. To this end, the intensity of an external control laser is varied only on a slow time scale, such that the system is kept in a coherent Raman dark state. When the control is ramped up, exactly one photon is emitted from the resonator. The electromagnetic field mode of the photon is determined by the properties of the control laser pulse \cite{morin_deterministic_2019}. Similarly, when the control field is ramped down, an impinging photon with matching temporal mode is absorbed \cite{boozer_reversible_2007}, and its polarization can be mapped to the spin state of the emitter \cite{specht_single-atom_2011}, as shown in Fig. \ref{fig_Raman}. This can be used to realize an efficient protocol for remote entanglement \cite{ritter_elementary_2012}. As the original atomic state is depleted, the scheme can be combined with state detection (detailed below) to herald successful entanglement attempts over a lossy channel.

\subsection{Spin initialization and readout} \label{sec_SpinReadout}

A key capacity for the processing of quantum information is the ability to perform a faithful projective measurement of the qubit state. Ideally, this readout procedure is robust and fast enough to allow for feedback onto the quantum state, which is a prerequisite for measurement-based quantum information processing \cite{briegel_measurement-based_2009}, entanglement distillation \cite{bennett_purification_1996, kalb_entanglement_2017}, and quantum error correction in the network \cite{nickerson_topological_2013}.

\begin{figure*}[t!] \centering
\includegraphics[width=2. \columnwidth]{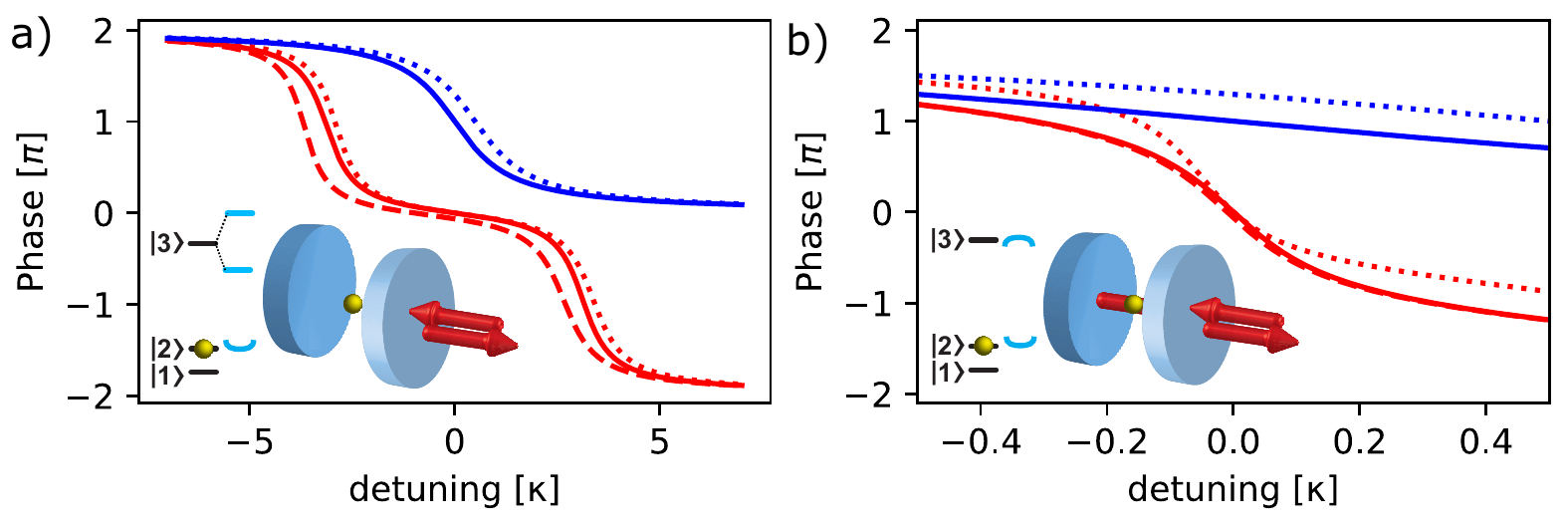}
\caption{\label{fig_GateMechanism}
\textbf{Cavity-based quantum gates. Insets:} Phase shift mechanism. A single photon ( arrows) is reflected from a single-sided optical resonator that contains a single emitter. The transition between the emitter energy levels (left) $\ket{2}$ and $\ket{3}$ is on resonance with the cavity frequency. When the emitter is in an off-resonant state $\ket{1}$, the light field enters the resonator before it is reflected, acquiring a phase shift of $\pi$. In the resonant emitter state $\ket{2}$, however, there is no phase shift of the combined emitter-photon state: a) In the strong-coupling regime the energy spectrum is split (dashed lines). Thus, the photon is reflected without entering the resonator, acquiring no phase shift. b) In the Purcell regime, the light field enters the cavity. It drives the emitter to the excited state from which it decays back into the cavity mode. In this process, both the emitter and the photon acquire a $\pi$ phase shift, such that their difference is again zero. \textbf{Main graphs:} Calculated phase difference as a function of the photon detuning (in units of the cavity linewidth $\kappa$) for systems with $C=50$ in the strong coupling regime (panel a, $g=\sqrt{10}\kappa > \kappa > \gamma = \kappa/10$) and in the Purcell regime (panel b, $\kappa > g=\kappa/\sqrt{10} > \gamma = \kappa/1000$). On resonance, a coupled emitter (red / gray) leads to a phase shift of $\pi$ with respect to the case of an empty cavity or an uncoupled emitter (blue / dark gray), almost independent of the emitter detuning (e.g. $\Delta_e=5\gamma$, dashed), but sensitive to the cavity detuning (e.g. $\Delta_c=0.5\kappa$, dotted).
}
\end{figure*}

With emitters in free space, the quantum state is typically measured via photon scattering on a closed transition \cite{leibfried_quantum_2003}. High readout fidelity is achieved when at least one scattered photon is detected before the spin decays to other levels via unwanted optical transitions or other decay mechanisms. This requires, first, frequency selective excitation of only one qubit state; second, a fast cycling transition that decays predominantly back to the original state; and third, highly efficient detectors and collection optics. Each of the above-mentioned criteria is improved when the emitter is placed in an overcoupled optical resonator that enhances the emission into a propagating light mode \cite{bochmann_lossless_2010}. When $P\gg 1$, single-shot readout can be achieved even with emitters that lack a closed transition \cite{raha_optical_2020, kindem_control_2020}.

However, the resonator also facilitates a different detection method, as the cavity transmission and reflection properties are altered by the presence of an emitter in a coupled energy level \cite{boozer_reversible_2007}. If the emitter and resonator frequency are known and stable, and in a regime where $C \gg 1$, this allows for state detection without photon scattering \cite{volz_measurement_2011}, which again means that the procedure works reliably for emitters that lack a closed transition. 

The described techniques leave the qubit in the measured quantum state, such that they can also be used for state initialization. However, often optical pumping is used to this end. Here, the idea is to repeatedly excite the emitter until it has decayed to the desired state, which should be the only level that is not pumped. The fidelity of the process depends on the ratio of the desired pumping rate versus that of off-resonant driving on unwanted transitions, and on the lifetime of the ground state. Again, a resonator can enhance this process by improving the frequency selectivity and speeding up the decay rate. Combining optical pumping with subsequent state detection may then provide an optimal initialization procedure in terms of speed and fidelity \cite{reiserer_nondestructive_2013}. 

\subsection{Spin-photon quantum gates} \label{sec_SpinPhotonGates}

The previous sections have described the generation and absorption of photons from a cavity-coupled emitter. But the resonator also enables another, deterministic interaction mechanism of coupled stationary qubits with impinging photons that are reflected from it \cite{hofmann_optimized_2003, duan_scalable_2004, borregaard_quantum_2019}. In particular, when $C \gg 1$, a spin-photon quantum gate can be realized without photon absorption or scattering. For an intuitive explanation of the mechanism, consider an emitter in a lossless, overcoupled cavity in the strong coupling regime, see Fig. \ref{fig_GateMechanism}a. A resonant photon is reflected off the coupling mirror which has a small transmission. If there is no emitter in the resonator, or the emitter is in an uncoupled qubit state, the light field leaking out of the resonator interferes destructively with the direct reflection at the coupling mirror, which means that the photon experiences a phase shift of $\pi$. If, however, a resonant emitter is present, the energy eigenstates of the coupled system are split \cite{reiserer_cavity-based_2015}. Thus, an impinging photon will now be off-resonant, meaning that it cannot enter the resonator but is reflected off the coupling mirror without a phase shift.

\begin{figure*}[t!] \centering
\includegraphics[width=2. \columnwidth]{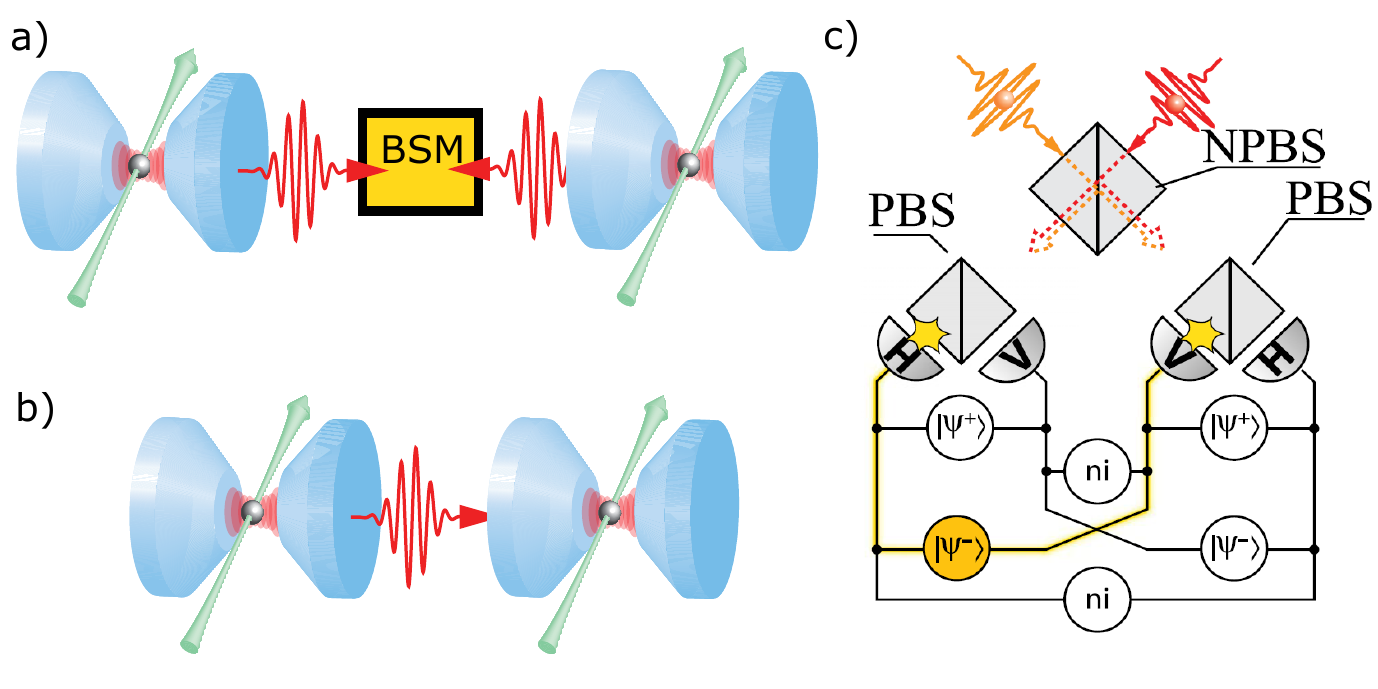}
\caption{\label{fig_RemoteEntanglement}
\textbf{Protocols to generate remote entanglement}. a) Entanglement swapping. First, an entangled spin-photon state is generated at both quantum network nodes. The photons impinge onto a setup that allows for a measurement of the photonic Bell state (BSM). b) Entanglement by heralded absorption. Spin-photon entanglement is generated at the left node. The photon is sent to the second node, where it is absorbed such that the encoded quantum state is transferred to the spin of the emitter. c) Linear optical Bell state measurement for polarization qubits. The photons impinge on a non-polarizing beam splitter (NPBS), followed by two polarizing beam splitters (PBS) with single photon detectors in their horizontal (H) and vertical (V) output ports. A coincidence detection in two of the output ports heralds a specific Bell state, $\ket{\Psi^+}$ or  $\ket{\Psi^-}$, depending on which detectors fire (yellow). In the other Bell states, photons have the same polarization an will thus leave the NPBS in the same output port. Therefore, coincidences with the same polarization indicate that the other photon properties were not identical (ni).
}
\end{figure*}

In effect, the reflection process leads to a conditional phase shift of $\pi$ between the emitter and the photon, i.e. a controlled-phase quantum gate. After first experiments with trapped atoms \cite{reiserer_nondestructive_2013, reiserer_quantum_2014, tiecke_nanophotonic_2014,  volz_nonlinear_2014}, quantum gates based on this mechanism have also been realized with superconducting qubits \cite{kono_quantum_2018}, quantum dots \cite{sun_single-photon_2018}, and spins in diamond \cite{nguyen_quantum_2019}. Remarkably, the fidelity of the scheme is robust to many experimental imperfections as long as the spatial mode and frequency of the photons match that of the emitter-cavity system. In particular, the magnitude of the phase shift does not depend on the precise emitter-cavity coupling strength and detuning, making the scheme well-suited for emitters with considerable spectral diffusion. The bandwidth of faithful operation is determined by the slope of the curves in Fig. \ref{fig_GateMechanism} - in the strong coupling regime, it is set by the cavity decay $\kappa$, whereas in the Purcell regime it depends on the enhanced emitter decay rate $g^2/\kappa$ \cite{kalb_heralded_2015}. 

For an ideal system, the scheme explained above is deterministic. In practice, the efficiency and fidelity can be reduced by imperfect optical mode matching, by cavity scattering loss, and by a finite cooperativity. The scaling with the latter has been the subject of several theoretical works that investigated entanglement generation or quantum gates based on cavity-induced phase shifts. Depending on the used protocol, one finds a scaling of the failure probability $\propto 1/\sqrt{C}$ \cite{sorensen_measurement_2003} or $\propto 1/C$ \cite{kastoryano_dissipative_2011}. Again, using the concept of heralding, one can achieve almost perfect fidelity as long as $C>1$, at the price of a success rate reduction that depends on $C$ \cite{borregaard_heralded_2015, borregaard_quantum_2019}.

The first experiment that implemented the above gate mechanism with trapped atoms at $C\simeq 3$ generated entangled states with $81\,\%$ fidelity \cite{reiserer_quantum_2014}, mainly limited by imperfect optical mode matching and single-qubit control. The current record has been achieved with the SiV center in diamond with $C \simeq 10^2$ and an entangled-state fidelity of $94\,\%$. While this number is encouraging for the implementation of quantum repeaters of the first generation \cite{briegel_quantum_1998, muralidharan_optimal_2016}, further improvements will be necessary for implementing one-way quantum repeaters that require much higher fidelity operations, e.g. $99.9\,\%$ in \cite{borregaard_one-way_2020}. Still, such schemes may eventually become feasible using cavity-coupled emitters with deterministic spin-photon coupling \cite{borregaard_quantum_2019}.

\subsection{Remote entanglement protocols} \label{sec_RemoteEntanglementProtocols}

As summarized in the above sections, optical resonators can be used to enhance the capabilities of quantum network nodes. In the following, I will describe the application of these techniques towards the generation of heralded entanglement between remote communication qubits, which is a key resource for quantum networks and required for first-generation quantum repeaters \cite{briegel_quantum_1998, muralidharan_optimal_2016}. In this context, two major approaches can be discriminated: First, entanglement swapping by photonic Bell-state measurements, and second, entanglement transfer by heralded absorption, as sketched in Fig. \ref{fig_RemoteEntanglement}.

\subsubsection{Entanglement swapping}

In the first approach, both quantum network nodes first generate spin-photon entanglement using the techniques described in Sec. \ref{sec_SpinPhotonEntanglement} and \ref{sec_RamanTransitions}. The two photons are then sent to a photonic Bell-state analyzer, which is typically realized by linear optical elements and single-photon detectors. Coincidence detection events then allow one to distinguish two out of four photonic Bell states \cite{calsamiglia_maximum_2001} which is sufficient for heralded remote entanglement via entanglement swapping \cite{zukowski_``event-ready-detectors_1993}. In principle, the intrinsic inefficiency can be avoided by photonic quantum gates that use deterministic schemes \cite{hacker_photonphoton_2016, stolz_quantum-logic_2021}, but so far do not reach the robustness, simplicity and efficiency of linear optical setups.

The physical effect that enables the Bell-state measurement in the latter is two-photon quantum interference \cite{hong_measurement_1987}. Note, however, that also single-photon interference protocols have been proposed \cite{cabrillo_creation_1999}, which may offer increased rates in a high-loss regime \cite{campbell_measurement-based_2008}, as successfully demonstrated in \cite{kalb_entanglement_2017}. In all such protocols, the mechanism only works if the photons are indistinguishable in all degrees of freedom except the one that encodes the entanglement with the spin \cite{reiserer_cavity-based_2015}. This enforces accurate control over the photon emission time, frequency, polarization, and wavepacket, which is difficult to achieve in practice, leading to a reduction in fidelity. Still, the latter may be improved at the price of a reduced success probability when the interference signal is recorded with high temporal resolution and only events that occur within a short arrival time difference are considered \cite{nolleke_efficient_2013, bernien_heralded_2013}. The above-mentioned approach also works in the absence of a cavity \cite{moehring_entanglement_2007, bernien_heralded_2013}, and reasonable efficiencies can be achieved in systems with strong optical transitions and optimized collection \cite{stephenson_high-rate_2020}. Still, embedding the emitter into a resonator can improve the rate and fidelity by enhancing the photon emission and collection probability, as detailed in Sec. \ref{sec_SpinPhotonEntanglement}.

\subsubsection{Heralded absorption}

A second approach to remote entanglement generation that is enabled or enhanced by optical resonators uses heralded photon absorption, which overcomes the efficiency limitation of photonic Bell state analyzers \cite{lutkenhaus_bell_1999} and can be more robust with respect to experimental imperfections. The process starts by spin-photon entanglement at one node. At the other node, the state of the photonic qubit is then transferred to the spin, as shown in Fig. \ref{fig_RemoteEntanglement}b. To allow for repeat-until-success entanglement, successful transfer has to be heralded.

Two options exist to achieve this task: First, the absorption process can be the time reversal of the photon emission when using stimulated Raman adiabatic passage \cite{cirac_quantum_1997} with multi-level emitters. This protocol can achieve the highest success probability and fidelity reported so far \cite{ritter_elementary_2012}. To herald successful qubit transfer, one has to detect whether the emitter has remained in its initial state or not, which can be accomplished using the techniques described in Sec. \ref{sec_SpinReadout}. Alternatively, the control laser field can be replaced by the vacuum field of a second resonator. Then, the detection of a scattered Raman photon signals qubit transfer to the spin \cite{brekenfeld_quantum_2020}.

The second method for heralded photon storage is based on the cavity-based spin-photon quantum gate mechanism described in Sec. \ref{sec_SpinPhotonGates}. The first realization used a combination of a gate operation with spin manipulations, photon detection, and active feedback \cite{kalb_heralded_2015}, but also a passive implementation has been demonstrated \cite{bechler_passive_2018}. Recently, the former approach has been used to generate entangled states by implementing a quantum gate between remote emitters \cite{daiss_quantum-logic_2021}, highlighting its potential for quantum networks.

\subsubsection{Ensemble-based approaches} \label{sec_ensembles}
The previous sections have focused on single emitters as quantum network nodes. This has the advantage that the non-linearity of the emitters allows for deterministic qubit interactions within a node, which improves the efficiency of first-generation and enables second- and third-generation quantum repeaters \cite{muralidharan_optimal_2016}. However, using single emitters requires resonators of very high quality. This difficulty is avoided when using the collective enhancement of the light-matter interaction with ensembles of emitters \cite{hammerer_quantum_2010}. Also in this approach, the exponential loss in optical fibers can be overcome with suited quantum repeater schemes, the first of which is often called the ``DLCZ protocol" \cite{duan_long-distance_2001}. In this scheme, atomic ensembles serve as both photon sources and quantum memories. Entanglement between remote ensembles is achieved in a repeat-until-success strategy by interfering emitted photons on a beam splitter. This induces a measurement-based nonlinearity, similar to that in related concepts for photonic quantum computing \cite{knill_scheme_2001}.

The main protocols and first experimental implementations of quantum networking with atomic ensembles have been summarized in \cite{sangouard_quantum_2011}. Early milestone experiments include the probabilistic entanglement \cite{chou_measurement-induced_2005} of up to four different ensembles \cite{choi_entanglement_2010} using atoms in vacuum. Later, also ensembles in solid-state platforms were employed to realize DLCZ-type photon sources \cite{laplane_multimode_2017, kutluer_solid-state_2017}, and the scheme has even been applied to entangle vibrations of remote optomechanical resonators \cite{riedinger_remote_2018}.

While the original DLCZ scheme uses ensembles both as quantum memory and entanglement source, the latter can also be implemented by other techniques, e.g. nonlinear optics. As an example, sources based on spontaneous parametric downconversion (SPDC) \cite{zhang_spontaneous_2021} can facilitate a speedup of the remote entanglement rate when they are combined with efficient and broadband quantum memories \cite{simon_quantum_2007}. Such devices can be realized with ensembles of trapped atoms or dopants in certain host crystals \cite{lvovsky_optical_2009, tittel_photon-echo_2010, afzelius_quantum_2015}, offering a large multiplexing capacity \cite{usmani_mapping_2010, seri_quantum_2017, afzelius_quantum_2015} that can be utilized in tailored quantum repeater protocols \cite{sinclair_spectral_2014} to facilitate high-rate remote entanglement. First experiments along these lines were the storage of entangled photons in two crystals \cite{clausen_quantum_2011, saglamyurek_broadband_2011}, also heralded after interfering photons at telecommunications wavelength \cite{lago-rivera_telecom-heralded_2021}. Other recent advances include the combination of an atomic ensemble photon source with a crystal-based memory, \cite{maring_photonic_2017}, and the entanglement of trapped-atom ensembles over $\SI{50}{\kilo\meter}$ of fiber \cite{yu_entanglement_2020}. The latter experiments used frequency conversion to the telecommunications frequency band, a prerequisite for entanglement over many kilometers of optical fibers.

While the above studies have been performed without optical resonators, their use can enhance the efficiency of both photon storage and -generation. When using an SPDC source, cavities can reduce the required optical driving power, give higher brightness of the source, improve the mode-matching to single-mode fiber, and facilitate spectral filtering, as summarized in \cite{slattery_background_2019}. But also when using emitter ensembles as photon source, their integration in optical resonators can lead to a high brightness \cite{thompson_high-brightness_2006} and enable a high readout efficiency of a stored excitation \cite{simon_interfacing_2007}. Furthermore, cavities can enhance other entanglement generation protocols, such as rephased amplified spontaneous emission \cite{williamson_cavity_2014}.

Also considering the storage of photons, cavities can boost the efficiency. The achievable enhancement can be understood semiclassically \cite{tanji-suzuki_chapter_2011} or treated in a quantum-mechanical framework \cite{gorshkov_photon_2007, afzelius_impedance-matched_2010}. Typical experiments use atoms in vacuum \cite{tanji_heralded_2009} or rare-earth-doped crystals \cite{sabooni_efficient_2013, jobez_cavity-enhanced_2014}. Remarkably, the enhancement of the light-matter-interaction strength offered by a cavity facilitates efficient quantum memories with a compact footprint, even down to nanophotonic devices \cite{zhong_nanophotonic_2017, wallucks_quantum_2020}. Furthermore, the use of optical resonators can suppress the noise in multiplexed quantum memories \cite{heller_cold-atom_2020}, enable couplings between different memory modes \cite{simon_single-photon_2007}, and realize additional functions such as light-pulse switching by stored photons \cite{chen_all-optical_2013} if $C>1$.

A drawback of the ensemble-based schemes presented above is the absence of an intrinsic nonlinearity. Thus, both in DLCZ and SPDC-based approaches the photon sources have to operate at a low efficiency to avoid the simultaneous emission of uncorrelated photons. Experimental imperfections lead to a further reduction of the remote entanglement rate, hampering large-scale quantum networks. To overcome this difficulty, adding a nonlinear processing capacity to resonator-enhanced ensembles seems attractive. The main approaches are based on the Coulomb interaction between trapped ions \cite{lamata_ion_2011, casabone_enhanced_2015}, and on Rydberg interactions in ensembles of neutral atoms. The latter facilitate interactions between photons \cite{firstenberg_attractive_2013}, and even photon-photon quantum gates in a free-space setting \cite{tiarks_photonphoton_2019}. Again, the use of optical resonators can dramatically enhance the efficiency of such approaches, with a recent experiment demonstrating $>40\,\%$ \cite{stolz_quantum-logic_2021}.

\section{Experimental realizations} \label{sec_experiments}

\subsection{Optical resonator designs}

\begin{figure*}[t!] \centering
\includegraphics[width=2. \columnwidth]{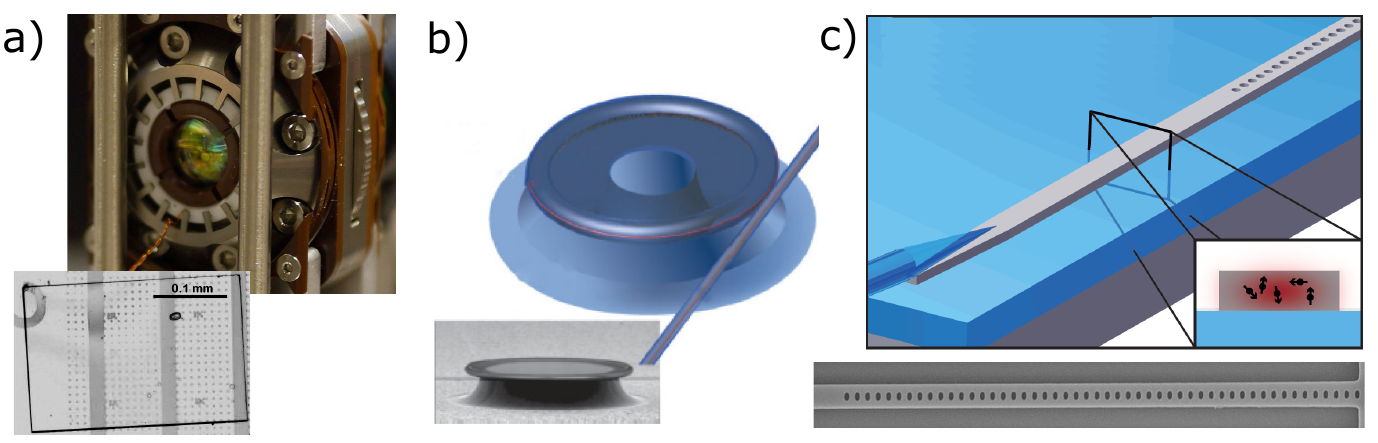}
\caption{\label{fig_ResonatorTypes}
\textbf{Types of optical resonators}. (a) Top: Experimental realization of a cryogenic Fabry-Perot cavity. The separation of the two mirrors (at the center), $<\SI{0.1}{\milli\meter}$, is controlled using a piezo tube (white ring) that presses against stiff titanium springs (gray), achieving length fluctuations $\lesssim 1\,\si{\pico\meter}$. The setup is mounted using soft polymer springs (dark, side of the assembly) that isolate it from environmental vibrations. From \cite{merkel_enhancing_2021}. Bottom: Microscope image of a diamond membrane with dimensions $0.01\times0.3\times0.2\,\si{\milli\meter}^3$, deposited on a Bragg reflector. Microwave striplines (dark gray) allow for the application of microwave pulses to emitters in the membrane. (b) Ring resonator in the form of a microtoroid that supports a whispering-gallery mode with high quality factor. The coupling to a close-by nanofiber can be adjusted via their distance. Bottom: Scanning electron microscope image of a resonator with a diameter of $120\,\si{\micro\meter}$. Adapted from \cite{vahala_optical_2003}. (c) Photonic crystal resonator. Top: Schematic of the setup. Using a conically tapered fiber, photons are coupled to a nanophotonic waveguide feeding the resonator at its end. Individual emitters are integrated into the resonator material, as seen in its cross section (inset). From \cite{weiss_erbium_2021}. Bottom: Scanning electron microscope image of a resonator, formed by a periodic arrangement of holes along a silicon waveguide with a separation of $\sim0.3\,\si{\micro\meter}$. 
}
\end{figure*}

The previous sections have focused on the underlying concepts of cavity-enhanced quantum network nodes. In the following, the current state-of-the-art of experimental systems will be summarized. An optical resonator that allows for the implementation of
an efficient quantum interface to single emitters should fulfill the following two conditions: $\kappa_\text{out} \gg \kappa_\text{loss}$ and $C \gg 1$. This indicates the requirement for resonators with small mode volume $V$ and large quality factor $Q$, as $C \propto Q/V$ \cite{reiserer_cavity-based_2015, lodahl_interfacing_2015, janitz_cavity_2020}. Several approaches exist towards realizing such a
resonator \cite{vahala_optical_2003}, the most prominent being Fabry-Perot, ring, and photonic crystal resonators, as shown in Fig. \ref{fig_ResonatorTypes}.

Fabry-Perot resonators consist of two curved mirrors at a short distance, as shown in Fig. \ref{fig_ResonatorTypes}a. In order to achieve high quality factors, Bragg reflectors are used, which consist of dielectric layers with alternating refractive indices, often $\text{Ta}_2\text{O}_5$ and $\text{SiO}_2$ with $n \simeq 2.1$ and $1.4$, respectively. The reflectors have to be deposited on atomically flat substrates to avoid excess loss by scattering. Transmission and scattering losses both below 1 ppm per mirror can be achieved with commercially available superpolished mirrors \cite{rempe_measurement_1992}, which gives a finesse around $\mathcal{F}=2\cdot10^6$, and even higher $Q=n\cdot\mathcal{F}$, where the mode number $n$ counts the half-waves in the resonant cavity. In typical experiments, cooperativities around 10 are achieved using this approach \cite{reiserer_cavity-based_2015}. Alternatively, low-roughness depressions with smaller radius of curvature can be fabricated by etching \cite{wachter_silicon_2019} or laser machining \cite{hunger_laser_2012}, both enabling finesse values beyond $2\cdot10^5$ and small mode volumes approaching a single cubic wavelength \cite{najer_gated_2019}.

Emitters can be integrated into the resonator either by trapping atoms in vacuum \cite{reiserer_cavity-based_2015}, or by depositing a nanocrystal \cite{kaupp_purcell-enhanced_2016, casabone_cavity-enhanced_2018} or a thin crystalline membrane \cite{janitz_fabry-perot_2015, bogdanovic_design_2017, riedel_deterministic_2017, merkel_coherent_2020, ulanowski_spectral_2021} on one of the mirrors. Experimentally achieved cooperativities \cite{colombe_strong_2007} and Purcell factors \cite{merkel_coherent_2020, ulanowski_spectral_2021} are of the order of $10^2$. Coupling to the resonator mode is achieved by free-space optics (with $>99\,\%$ efficiency) or by directly coupling to a single-mode fiber with $\sim90\,\%$ efficiency \cite{gulati_fiber_2017, niemietz_nondestructive_2021}.

Compared to the other approaches described below, Fabry-Perot resonators have two major advantages: First, they can be stabilized and tuned over many free spectral ranges using piezoelectric positioners. Second, to first order, the cooperativity does not depend on the cavity length $L$, as $Q\propto L$ and $V\propto L$. Thus, without reduction of the cooperativity an emitter can be kept at a large distance from all interfaces, which avoids the undesired influence of surface charges and paramagnetic trap states on the emitter stability.

A second approach to implement resonators with large cooperativity uses ring resonators, as shown in Fig. \ref{fig_ResonatorTypes}b, either based on a whispering-gallery mode in microtoroids \cite{aoki_observation_2006}, microspheres \cite{shomroni_all-optical_2014}, or bottle resonators \cite{pollinger_ultrahigh-q_2009}, or using nanophotonic waveguide ring or racetrack resonators \cite{bogaerts_silicon_2012}. Emitters in the mode can exhibit a chiral coupling to light, leading to new possibilities for spin-photon interfaces \cite{lodahl_chiral_2017}. Tuning is typically achieved by temperature \cite{aoki_observation_2006} or, with bottle resonators, mechanically \cite{pollinger_ultrahigh-q_2009}. Experimentally achieved cooperativities with atoms \cite{aoki_observation_2006, junge_strong_2013, scheucher_quantum_2016, bechler_passive_2018} and Purcell factors with defect centers \cite{faraon_resonant_2011} are on the order of $10$, and high coupling efficiency is obtained via tapered fibers.

The third approach for efficient light-matter coupling is based on photonic crystal resonators \cite{lodahl_interfacing_2015, asano_photonic_2018}, as shown in Fig. \ref{fig_ResonatorTypes}c. When coupled to single emitters, cooperativities approaching $10^2$ \cite{samutpraphoot_strong_2020} and Purcell factors approaching $10^3$ \cite{dibos_atomic_2018} have been reported. Optimized structures in silicon even enable $Q>10^7$ \cite{asano_photonic_2017} at mode volumes around $\lambda^3$.  When using dielectric enhancement, the effective mode volume can be further reduced by three orders of magnitude \cite{hu_experimental_2018} while maintaining high Q. To use such a structure for quantum network nodes, however, the communication qubits have to be placed in the dielectric material of the resonator, which limits the applicability of the approach to specific combinations of emitter and host. Furthermore, the close proximity of interfaces will likely degrade the coherence of the emitter in such setting. Finally, care has to be taken in the evaluation of the cooperativity, as the dipole approximation assumes that the electric field only changes on a scale that is comparable to the wavelength \cite{cohen-tannoudji_photons_1989}, which is not satisfied in structures with deeply subwavelength dielectric features.

Tuning of photonic crystal cavities has been demonstrated by many techniques, including gas condensation (in the case of cryogenic resonators) \cite{mosor_scanning_2005}, nanomechanical actuation \cite{chew_dynamic_2010}, electro-optical shifting \cite{lu_lithium_2012} and temperature \cite{tiecke_nanophotonic_2014}. To couple into the resonators, different techniques can be used. The highest efficiencies, $97\,\%$, are achieved using an adiabatic transition of the guided mode of a tapered optical fiber to that of a high-index dielectric waveguide \cite{tiecke_efficient_2015} feeding the cavity. Other approaches use cleaved or lensed fibers with mode converters at the chip edge or diffraction gratings at the chip center, with typical efficiencies of $50\,\%$ \cite{vivien_handbook_2013}.

\subsection{Experimental platforms}

\begin{figure*}[t!]\centering
\includegraphics[width=2. \columnwidth]{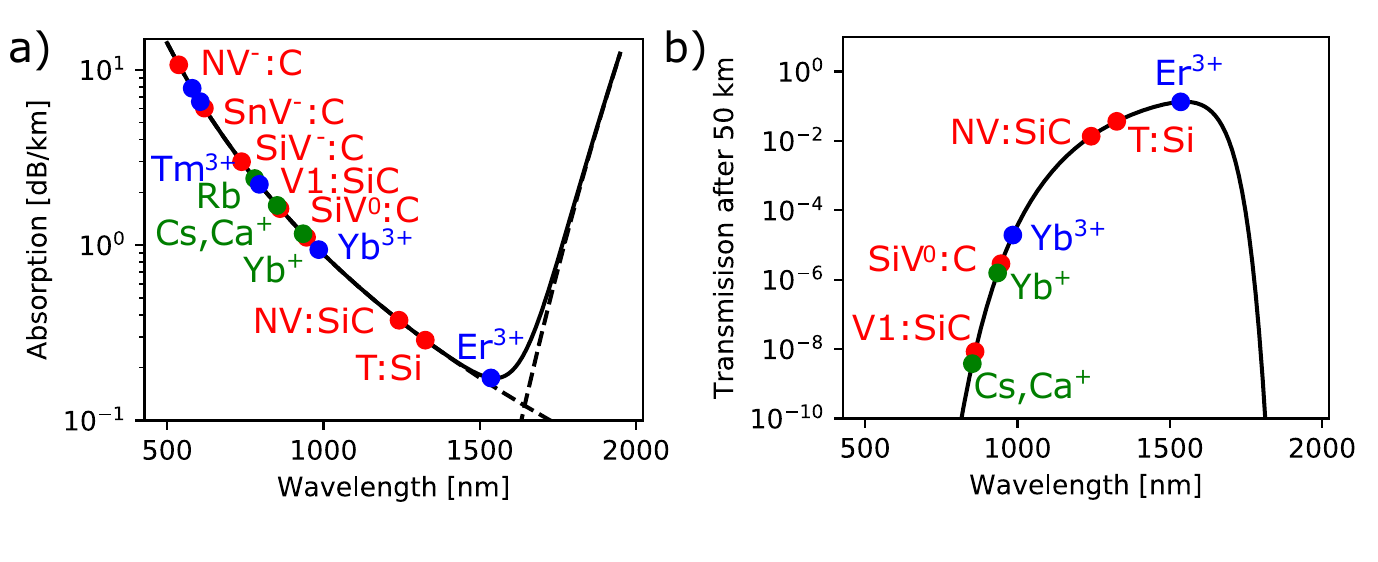}
\caption{\label{fig_fiberloss}
\textbf{Loss in optical fibers}. a) Absorption coefficient (black solid) of ultrapure "dry" silica fiber caused by Rayleigh scattering and infrared absorption (black dashed lines). Only few emitters (defects in diamond [C] and other semiconductors [Si and SiC]; atoms in vacuum [Rb, Cs, $\text{Ca}^{+}$, $\text{Yb}^{+}$]; rare-earth dopants [$\text{Tm}^{3+}$, $\text{Yb}^{3+}$, $\text{Er}^{3+}$]) fall in the low-loss telecommunications window  between $~1250\,\si{\nano\meter}$ and $1650\,\si{\nano\meter}$. b) Transmission after $50\,\si{\kilo\meter}$ of optical fiber. At visible wavelengths, losses seem prohibitive. In contrast, in the telecommunications window the $10\,\%$ transmission may be sufficient for quantum networking at a reasonable rate.}
\end{figure*}

In the following section, the physical systems that are used as communication qubits are described. As mentioned, their main purpose is to provide an efficient interface to photons, which is achieved by a suited optical resonator. Ideally, the photon wavelength will fall in the so-called telecommunications window, between $1500$ and $1600\,\si{\nano\meter}$, where the loss of germanium-doped silica optical fibers is minimal  \cite{lines_search_1984}, around $0.2\,\text{dB/km}$, see Fig. \ref{fig_fiberloss}a. While optical fiber links with lower loss would be desirable for global networks, no such system has been demonstrated in spite of an intense search over several decades. Still, when using the existing infrastructure photonic qubits can be transmitted over many kilometers with negligible decoherence even at room temperature and with moderate loss, see Fig. \ref{fig_fiberloss}b.

To couple to these photons, the most prominent physical systems explored so far are single atoms (green), impurities in diamond, silicon or silicon carbide (red), and rare-earth dopants (blue). Quantum dots with their wide tunability are not included in the figure as it seems difficult to combine them with memory qubits that offer sufficient coherence for long-distance networks \cite{lodahl_interfacing_2015}. Furthermore, the figure only contains transitions that have been investigated in experiments, which all originate from a long-lived ground state. It has been proposed that transitions in the excited state manifold of neutral atoms may offer telecom-compatibility when resonators with large cooperativity are used \cite{uphoff_integrated_2016, covey_telecom-band_2019, menon_nanophotonic_2020}, but an experimental demonstration is still missing.

Fig. \ref{fig_fiberloss}b thus indicates that most investigated systems will require efficient transduction of the photon frequency \cite{zaske_visible--telecom_2012} when large distances are targeted. This will add complexity and cost while reducing the efficiency and fidelity. Still, recent advances have enabled entanglement-preserving telecom conversion from visible emitters \cite{de_greve_quantum-dot_2012, bock_high-fidelity_2018, tchebotareva_entanglement_2019, van_leent_long-distance_2020}, demonstrating the feasibility of the approach.

As an alternative, emitters in the minimal loss band of optical fibers, i.e. between $\sim\SI{1250}{\nano\meter}$ and  $\sim \SI{1650}{\nano\meter}$, can be used. These include defect centers in silicon \cite{bergeron_silicon-integrated_2020, durand_broad_2021} and silicon carbide \cite{wang_coherent_2020} (with so-far unknown optical coherence), as well as erbium dopants. The optical transitions of the latter can exhibit remarkable coherence of several ms in some host materials \cite{bottger_optical_2006}, approaching the lifetime limit in suited resonators \cite{merkel_coherent_2020, ulanowski_spectral_2021}. However, owing to the ms-long lifetime of their telecom transition in all studied hosts \cite{bottger_optical_2006, stevenson_erbium-implanted_2021, weiss_erbium_2021, gritsch_narrow_2021}, using single dopants in this platform requires resonators with large Purcell enhancement factors, which have only been demonstrated recently \cite{dibos_atomic_2018, merkel_coherent_2020}.

Entanglement generation between remote erbium dopants is still an outstanding challenge. The current state in this respect will be presented in Sec. \ref{sec_rare_earth_dopants}. In contrast, elementary quantum network links have been achieved in several other platforms. Most notably, remote entanglement generation has been demonstrated with $\text{Yb}^+$ \cite{moehring_entanglement_2007, hucul_modular_2015} and $\text{Sr}^+$ \cite{stephenson_high-rate_2020} ions, Rb atoms \cite{hofmann_heralded_2012, ritter_elementary_2012}, and NV centers in diamond \cite{bernien_heralded_2013}. Therefore, these hardware platforms will be explained in detail in the following sections.

\begin{figure*}[t!]\centering
\includegraphics[width=2. \columnwidth]{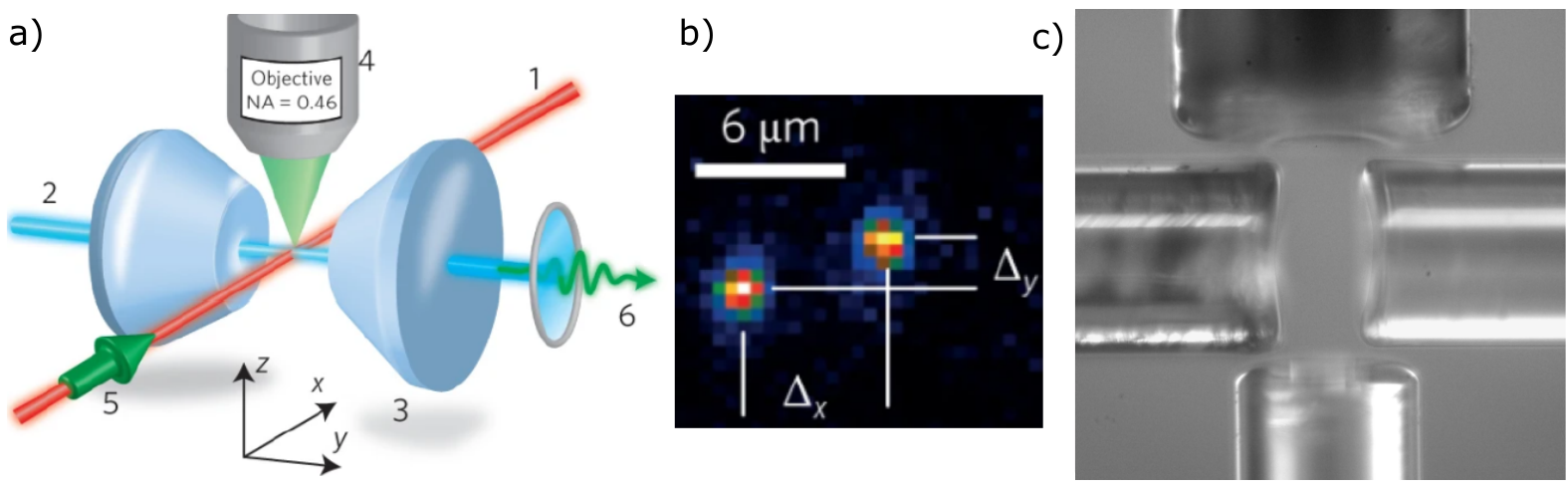}
\caption{\label{fig_TrappedAtom}
\textbf{Trapped-atom quantum network nodes.} a) Schematic of a typical setup. Atoms are trapped in a Fabry-Perot resonator (3) using standing-wave laser fields (1 and 2), and an objective (4) collects scattered light for imaging. b) Fluorescence image used to determine the number of loaded atoms and their position along $x$ and $z$. Individual addressing is possible with tightly-focused laser beams in order to realize a quantum network node with several stationary qubits. Adapted from \cite{neuzner_interference_2016}. c) Photograph of a crossed-cavity setup. Each of the two resonators consists of two coated glass fiber end facets with Gaussian depressions generated by laser ablation. A single atom can be coupled simultaneously to both resonators, which facilitates advanced quantum networking protocols. Picture of the setup of \cite{brekenfeld_quantum_2020}.
}
\end{figure*}

\subsubsection{Atoms in vacuum} \label{sec_atoms}

Many pioneering experiments in the field of quantum networks have used atoms trapped in vacuum. Since the first generation of remote entanglement \cite{moehring_entanglement_2007}, several other experiments have achieved this milestone \cite{ritter_elementary_2012, hofmann_heralded_2012, hucul_modular_2015, stephenson_high-rate_2020, daiss_quantum-logic_2021}. By integrating the atoms into optical resonators, high efficiencies and many advanced protocols have been realized \cite{reiserer_cavity-based_2015}, including teleportation \cite{nolleke_efficient_2013, langenfeld_quantum_2021}, quantum memories with single \cite{specht_single-atom_2011, kalb_heralded_2015, brekenfeld_quantum_2020} and several \cite{casabone_enhanced_2015, langenfeld_network-ready_2020} atoms, photon-mediated quantum gates \cite{reiserer_quantum_2014, tiecke_nanophotonic_2014, hacker_photonphoton_2016, daiss_quantum-logic_2021, dordevic_entanglement_2021}, nondestructive photon \cite{reiserer_nondestructive_2013, distante_detecting_2021} and photonic qubit \cite{ niemietz_nondestructive_2021} detection, and basic quantum repeater nodes \cite{langenfeld_quantum_2021-1}. These advances have established trapped atoms as one of the leading experimental platforms for quantum networks.

To use atoms in vacuum as stationary and efficient network nodes, they have to be localized to a subwavelength spot. To this end, typically tight trapping potentials with trap frequencies on the order of a few hundred kHz are employed. The atoms are then confined in the Lamb-Dicke regime \cite{leibfried_quantum_2003}, where the motional state of the atom only occasionally changes in absorption and emission events. Still, efficient laser re-cooling is possible by various techniques \cite{reiserer_cavity-based_2015}, leaving the atom in the
ground state of the potential \cite{reiserer_ground-state_2013}.

To implement the required trap, two approaches can be followed: First, electrical trapping of charged atoms \cite{leibfried_quantum_2003}, and second, optical trapping in far-detuned laser fields \cite{grimm_optical_2000}. Importantly, both traps can be integrated with optical resonators in order to enhance the efficiency of spin-photon interactions, as shown in Fig. \ref{fig_TrappedAtom}. As several atoms can be loaded to the same trap, quantum network nodes with several qubits can be realized \cite{casabone_heralded_2013, neuzner_interference_2016}. These qubits can be different atomic species to avoid cross-talk during optical addressing and control \cite{inlek_multispecies_2017}.

When trapped in vacuum, atoms are well isolated both from the environment and from one another. Thus, they can exhibit very long coherence times. With neutral atoms in optical resonators, encoding the qubit in a magnetic-field insensitive state has enabled a spin-echo time exceeding $100\,\si{\milli\second}$ \cite{korber_decoherence-protected_2018}, which is already promising for extended quantum networks. Eventually, in deep optical dipole traps the coherence will be limited by scattering of trap photons and by the requirement to periodically recool the atoms. This is avoided in electrical traps, where sympathetic cooling has recently enabled coherence times on the scale of one hour for $\text{Yb}^+$, still far from the fundamental limitations of background gas scattering and hyperfine lifetime \cite{wang_single_2021}.

Such long coherence times pose no restrictions to the fidelity of single- and two-qubit operations within a node. Also, most technical limitations can be avoided by careful experimental design, by advanced pulses and pulse sequences adapted from nuclear magnetic resonance \cite{vandersypen_nmr_2005}, and by optimal control theory \cite{werschnik_quantum_2007}. In this way, very high fidelities -- exceeding $99.99\,\%$ (in the absence of a resonator) \cite{ballance_high-fidelity_2016} -- for the preparation of arbitrary single qubit states have been demonstrated. To this end, the atom is first initialized to a single state by optical pumping. Then, irradiation of electromagnetic fields at the frequency of the qubit transition can induce arbitrary rotations. The use of optical rather than microwave fields eases individual addressing of several qubits in the same trap. A more detailed description of the techniques for single-atom control is given e.g. in \cite{reiserer_cavity-based_2015}.

In addition to the high-fidelity initialization, also faithful readout of the atomic state can be achieved using fluorescence state detection both in free-space and in optical resonators. In the latter, as described in Sec. \ref{sec_SpinReadout} also resonator transmission can be used to reduce or even avoid readout-induced heating by photon scattering \cite{volz_measurement_2011}.

The implementation of quantum repeaters with dedicated memory and communication qubits also requires to control several atoms \cite{casabone_enhanced_2015, neuzner_interference_2016} with individual addressing \cite{langenfeld_network-ready_2020, langenfeld_quantum_2021} and local deterministic two-qubit operations. To date, high-fidelity gates based on the Coulomb interaction have been achieved even between ions of different species \cite{negnevitsky_repeated_2018, hughes_benchmarking_2020}, with fidelities of $99.8\,\%$ in the absence of a resonator. For neutral atoms, gates can be implemented via photonic interactions \cite{welte_photon-mediated_2018}, or via dipolar coupling in a highly excited Rydberg state \cite{saffman_quantum_2010}, enabling entanglement generation fidelities $>99\,\%$ in the absence of a resonator \cite{madjarov_high-fidelity_2020}. 
Finally, the realization of quantum networks requires entanglement between remote nodes with high success probability $\eta$ and fidelity $F$. In free space, values of $\eta=2\cdot10^{-4}$ and $F=94\,\%$ have been demonstrated with trapped ions \cite{stephenson_high-rate_2020}. Neutral atoms in optical resonators have achieved $\eta=2\,\%$ and $F=85\,\%$ based on a Raman absorption protocol \cite{ritter_elementary_2012}, and $\eta=0.6\,\%$ and $F=79\,\%$ based on a remote quantum gate protocol \cite{daiss_quantum-logic_2021}. As no fundamental limitations have been identified in these experiments, it seems likely that these values can be further improved, either based on the previously used protocols or on novel approaches towards heralded qubit storage \cite{bechler_passive_2018, brekenfeld_quantum_2020}. Eventually, exceeding the error correction threshold of surface codes \cite{fowler_surface_2012} in a networked topology \cite{nickerson_topological_2013} ($F \gtrapprox 90\,\%$ for remote entanglement) seams feasible. In this context, the $F=98\,\%$ achieved when post-selecting on correct atomic state initialization \cite{ritter_elementary_2012} is encouraging.

To summarize, trapped atoms are a leading platform for the implementation of quantum networks and repeaters. The next steps towards the latter will likely involve the development of systems with more individually controlled qubits per node \cite{hucul_modular_2015, casabone_enhanced_2015,  langenfeld_network-ready_2020, langenfeld_quantum_2021}, potentially with dedicated communication and memory qubits. Compared to other platforms under study, the main advantage of atoms trapped in vacuum is their excellent isolation, which enables long coherence and operations with exceptional fidelity. However, this comes at the price of requiring ultra-high vacuum and advanced optical setups with precisely stabilized high-power lasers. Albeit such systems have been realized in many laboratories, the implementation of field-deployable quantum network nodes based on trapped atoms is an outstanding engineering challenge.

\subsubsection{Defect centers in semiconductors} \label{sec_defects}

Because of the mentioned technical overhead required to trap and cool single atoms in vacuum, significant effort has been invested in the search for solid-state alternatives. The first such system that has received considerable attention is the nitrogen-vacancy (NV) center in diamond. Landmark experiments with this platform include the demonstration of spin-photon entanglement \cite{togan_quantum_2010}, remote entanglement \cite{bernien_heralded_2013} over distances up to $1.3\,\si{\kilo\meter}$ \cite{hensen_loophole-free_2015}, the unconditional teleportation of a quantum state \cite{pfaff_unconditional_2014}, the distillation of entanglement between remote quantum network nodes \cite{kalb_entanglement_2017}, the deterministic delivery of remote entanglement \cite{humphreys_deterministic_2018}, and the realization of a three-node quantum network \cite{pompili_realization_2021}.

These experiments have been facilitated by the remarkable coherence properties of NV center spins in diamond \cite{doherty_nitrogen-vacancy_2013} up to room temperature, which forms the basis for many applications in quantum sensing \cite{degen_quantum_2017}. When transferring qubits to the nuclear spin of close-by $^{13}\text{C}$ atoms, coherence can even be preserved for seconds \cite{maurer_room-temperature_2012}. Unfortunately, the coupling to phonons prevents the use of NV centers for remote entanglement generation at room temperature, as it leads to fast mixing of the excited state spin \cite{doherty_nitrogen-vacancy_2013}. Instead, cryogenic operation at a typical temperature of $4\,\si{\kelvin}$ is required to this end.

At such temperature, the individual optical transitions of the NV center can be resolved \cite{batalov_low_2009}. Some of them preserve the spin state well \cite{tamarat_spin-flip_2008} and can thus be used for single-shot readout with high fidelity, provided the photon collection efficiency is high. This has first been achieved  \cite{robledo_high-fidelity_2011} by placing the NV center into a solid immersion lens \cite{hadden_strongly_2010}. In this way, total internal reflection in high-refractive index host materials is avoided. The solid immersion lens can be combined with antireflective coatings to enhance the collection efficiency towards the theoretical maximum of $50\,\%$ when using a lens system with a high numerical aperture.

Efficient collection also helps to increase the rate of remote entanglement by two-photon interference \cite{bernien_two-photon_2012, sipahigil_quantum_2012}, which typically starts with the generation of spin-photon entanglement at both remote nodes. While initial experiments used polarization qubits \cite{togan_quantum_2010}, an alternative scheme based on time-bin qubits \cite{barrett_efficient_2005} turned out to be more robust \cite{bernien_heralded_2013}, as it comes naturally with the decoupling of magnetic dephasing.

\begin{figure}[t!]\centering
\includegraphics[width=1. \columnwidth]{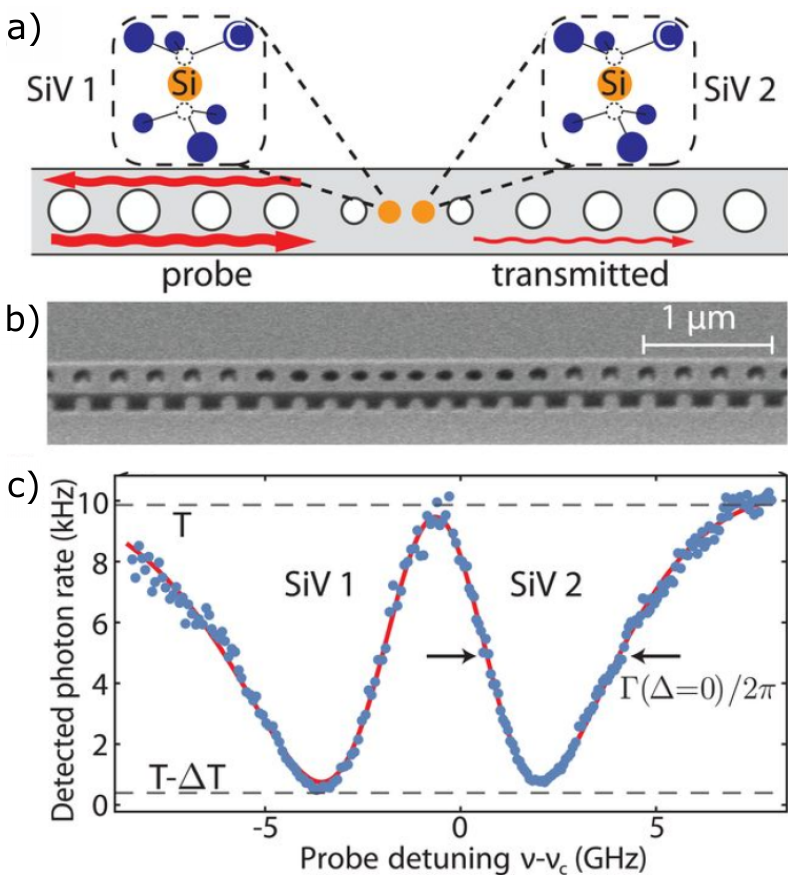}
\caption{\label{fig_SiV}
\textbf{Quantum network node based on defect centers in diamond.} a) Typical experimental setting. Light is confined along the direction of a waveguide with triangular cross-section using a periodic pattern of holes. Individual defect centers, here two SiV-centers, are generated at the field maximum by implantation and annealing. Light is coupled to the resonator via a tapered optical fiber attached to a tapered end of the waveguide (not shown). b) Scanning-electron-micrograph of a diamond resonator, fabricated by reactive ion etching. c) Spectral signature of the SiV-resonator coupling. The cavity is tuned such that the probe laser beam is on resonance. At the transition frequencies of two coupled SiVs, the transmission is almost completely suppressed, testifying the good emitter-resonator coupling, which can be quantified from the broadening of the SiV transition linewidth $\Gamma$. Adapted from \cite{evans_photon-mediated_2018}.
}
\end{figure}

To realize this scheme and prove entanglement by reading the spin state in different bases, ground state control needs to be implemented. Albeit all-optical control can also be used with defect qubits \cite{santori_coherent_2006, yale_all-optical_2013}, providing minimal crosstalk in dense systems, experiments typically use microwave pulses for their ease of implementation. In the NV center, the combination of a small magnetic bias field with the zero-field splitting of the defect \cite{doherty_nitrogen-vacancy_2013} leads to transition frequencies around $3\,\si{\giga\hertz}$, which can be conveniently applied via close-by wires or microwave striplines. In this way, control fidelities around $99.9\,\%$ are routinely achieved \cite{hensen_loophole-free_2015}. Such high pulse fidelities also allow for dynamical decoupling with many control pulses to extend the coherence time of the electronic spin beyond $1\,\si{\second}$ at cryogenic temperature.

This is possible even in samples with natural isotope abundance \cite{abobeih_one-second_2018}, where about $99\,\%$ of the carbon atoms have no nuclear spin. The few remaining $^{13}\text{C}$ spins in the proximity of the NV electronic spin can be used as an additional resource for quantum information processing. In particular, they can serve as quantum registers \cite{dutt_quantum_2007, neumann_quantum_2010}, potentially with error correction \cite{waldherr_quantum_2014, cramer_repeated_2016}, and as robust memory qubits in a quantum network node \cite{reiserer_robust_2016}, as they are decoupled from the optical channels and only interact with adjacent spins. Thus, using tailored sequences, the nuclear spin state is preserved for thousands of entanglement attempts \cite{kalb_dephasing_2018}. In combination with the recently demonstrated potential for minute-long natural dephasing times \cite{bartling_coherence_2021}, this makes nuclear spin registers a unique resource for quantum networks.

To harness the potential of nuclear spins, they have to be controlled with high fidelity via the hyperfine interaction with the electronic spin. Strongly coupled spins can be controlled via frequency-selective electromagnetic fields \cite{jelezko_observation_2004, dutt_quantum_2007}, but quickly lose their coherence when the electronic spin undergoes a random flip during entanglement generation attempts \cite{blok_towards_2015}. Therefore, the control of spin registers with weaker coupling is preferable. This comes at the price of slower local operations, whose speed is, however, not limiting in typical long-distance experiments. The required universal control can be achieved with a sequence of microwave pulses \cite{taminiau_universal_2014}, potentially in combination with radiofrequency pulses to enhance the number of controllable spins and further improve the control fidelity \cite{bradley_ten-qubit_2019}.

The main challenge towards the use of the NV center in quantum network nodes is the inefficiency of its zero-phonon optical transition \cite{doherty_nitrogen-vacancy_2013}. Albeit single-photon protocols can substantially improve the rate \cite{campbell_measurement-based_2008, kalb_entanglement_2017} and thus facilitate repeat-until-success entanglement \cite{humphreys_deterministic_2018}, for large-distance experiments the achievable rates seem prohibitively low. Therefore, it would be desirable to enhance the emission into the zero phonon line via the Purcell effect. This has first been achieved in nanophotonic resonators \cite{faraon_resonant_2011}, but the spectral diffusion of the optical transition observed in these experiments has hindered remote entanglement.

The frequency instability is attributed to charge fluctuations. While the state of the NV center itself can be well controlled \cite{siyushev_optically_2013, doi_deterministic_2014}, the Stark effects induces large jumps of the optical transition frequency when changing the state of close-by charge traps. In pure bulk crystals, the effect is small enough to be compensated by feedback \cite{robledo_high-fidelity_2011, acosta_dynamic_2012}, but the proximity of charge traps at the interface impedes the use of nanostructured diamond resonators.

A possible solution is to integrate bulk crystals with embedded NVs into Fabry-Perot resonators with small mode volume and high finesse. Purcell enhancement has also been demonstrated in such setting \cite{riedel_deterministic_2017}. Still, in spite of recent progress \cite{casabone_dynamic_2021, fontana_mechanically_2021}, it has turned out difficult to achieve the required length stability of a cavity with transversal scanning ability when operating in closed-cycle cryogenic systems \cite{janitz_fabry-perot_2015,bogdanovic_design_2017} with their typically strong vibrations. Instead, positioning individual defects within a rigid tube resonator assembly, as demonstrated with rare-earth dopants \cite{merkel_coherent_2020}, may provide a viable solution.

An alternative to such efforts is to use different defects than the NV center. In particular, the absence of a linear Stark shift for defects with inversion symmetry \cite{macfarlane_optical_2007} is beneficial for quantum networks nodes. Pioneering work used the SiV center in diamond that has promising optical properties in the negatively charged \cite{hepp_electronic_2014, rogers_all-optical_2014} and neutral \cite{rose_observation_2018} state, which both show stable transition frequencies and a comparably large fraction of zero-phonon-line emission. Two-photon interference experiments have demonstrated good photon indistinguishability \cite{sipahigil_indistinguishable_2014}, which formed the basis for spin-spin entanglement by detecting photons emitted into a waveguide \cite{sipahigil_integrated_2016}. Experiments with photonic crystal resonators, as shown in Fig. \ref{fig_SiV}, have now paved the way towards entanglement and quantum networking experiments \cite{evans_photon-mediated_2018, nguyen_quantum_2019} based on the phase-shift mechanism presented in Sec. \ref{fig_GateMechanism}. Remarkably, memory-enhanced quantum communication \cite{bhaskar_experimental_2020} and the entanglement of several frequency-multiplexed emitters in the same resonator \cite{levonian_optical_2021} have been demonstrated in this platform, demonstrating the key steps required for implementing a quantum repeater, c.f. Fig \ref{fig_repeater}.

Still, a drawback of the negatively charged SiV center is that so far, sufficient coherence of the ground state has only been obtained at $\si{\milli\kelvin}$ temperature \cite{jahnke_electronphonon_2015, sukachev_silicon-vacancy_2017, becker_all-optical_2018}. Therefore, other group-IV defects that may operate at higher temperature, such as the neutral SiV \cite{rose_observation_2018} and the SnV \cite{trusheim_transform-limited_2020, rugar_quantum_2021} may be favorable to enhance the prospects towards upscaling. In this respect, the difficulty to grow pure diamond samples on a wafer scale -- albeit favorable for the jewelry industry -- may be an obstacle unless hybrid integration is used \cite{wan_large-scale_2020}. 
This challenge is less pronounced in other large-bandgap semiconductors, such as silicon carbide. Also in this material, a large number of defects with promising properties have been identified. A recent overview is given in \cite{atature_material_2018, wolfowicz_quantum_2021}. In particular, silicon vacancy centers \cite{riedel_resonant_2012, nagy_high-fidelity_2019} have demonstrated the generation of indistinguishable photons \cite{morioka_spin-controlled_2020} with high efficiency \cite{lukin_integrated_2020, babin_nanofabricated_2021}, making SiC a promising candidate for the scaling of quantum networks. Note, however, that most defects in SiC and diamond emit light at frequencies where the transmission of optical fibers is moderate, as shown in Fig. \ref{fig_fiberloss}. Therefore, photon conversion to the telecom band will be required to bridge global distances. Still, there are several emitters in the O-band around $1300\,\si{\nano\meter}$ in SiC \cite{wang_coherent_2020, wolfowicz_vanadium_2020} and silicon \cite{bergeron_silicon-integrated_2020, durand_broad_2021} that may be used over larger distances without wavelength conversion.

\subsubsection{Rare-earth dopants} \label{sec_rare_earth_dopants}

In spite of the well-developed quantum network nodes based on trapped atoms and defects in large-bandgap semiconductors, the search for qubit systems with improved properties has not come to an end. In recent years, a third promising platform has emerged in this context: Crystals with rare-earth dopants, typically in the triply ionized state. These emitters exhibit optical transitions between electronic states in the inner 4f shell \cite{thiel_rare-earth-doped_2011}, which are surrounded by filled 5s and 5p shells. The electrons in these outer orbitals shield the electric field of neighboring atoms in the crystal to a remarkable degree. Thus, the crystal field can be treated as a small perturbation to the energy levels of the free ion \cite{liu_spectroscopic_2005}, such that the optical transition frequencies are almost independent of the host crystal. Remarkably, for one of the rare-earth dopants, erbium, these transitions fall within the telecommunications window around $1550\,\si{\nano\meter}$. This is not only the basis for erbium-doped fiber lasers and amplifiers that are widespread in classical networks, but also makes this emitter an interesting candidate for quantum networks. In this context, the coherence of the optical transitions is paramount. Because of the shielding effect, at cryogenic temperature the coupling to phonons plays a negligible role in most hosts, and optical coherence of several $\si{\milli\second}$ is obtained in some systems \cite{bottger_optical_2006} --- the longest observed in any solid.

Decoherence rates can also be extremely low in the ground state, where the precise value depends on the dopant, host crystal, and magnetic field \cite{thiel_rare-earth-doped_2011}. In some systems, spin lifetimes of several weeks are observed, which forms the basis for the realization of quantum memories with exceptional lifetime of several hours \cite{zhong_optically_2015}. In this context, the ideal host crystal exhibits a high Debye temperature, has a large bandgap and a low impurity concentration, and is free of nuclear magnetic moments \cite{atature_material_2018,zhong_emerging_2019, wolfowicz_quantum_2021}. In addition, the dopants should be integrated at a well-defined lattice site without generating too much strain or fluctuating charge traps. Finally, the crystal field levels should be well-split, which reduces phononic relaxation at a given temperature \cite{liu_spectroscopic_2005, wolfowicz_quantum_2021}.

\begin{figure*}[t!]\centering
\includegraphics[width=2. \columnwidth]{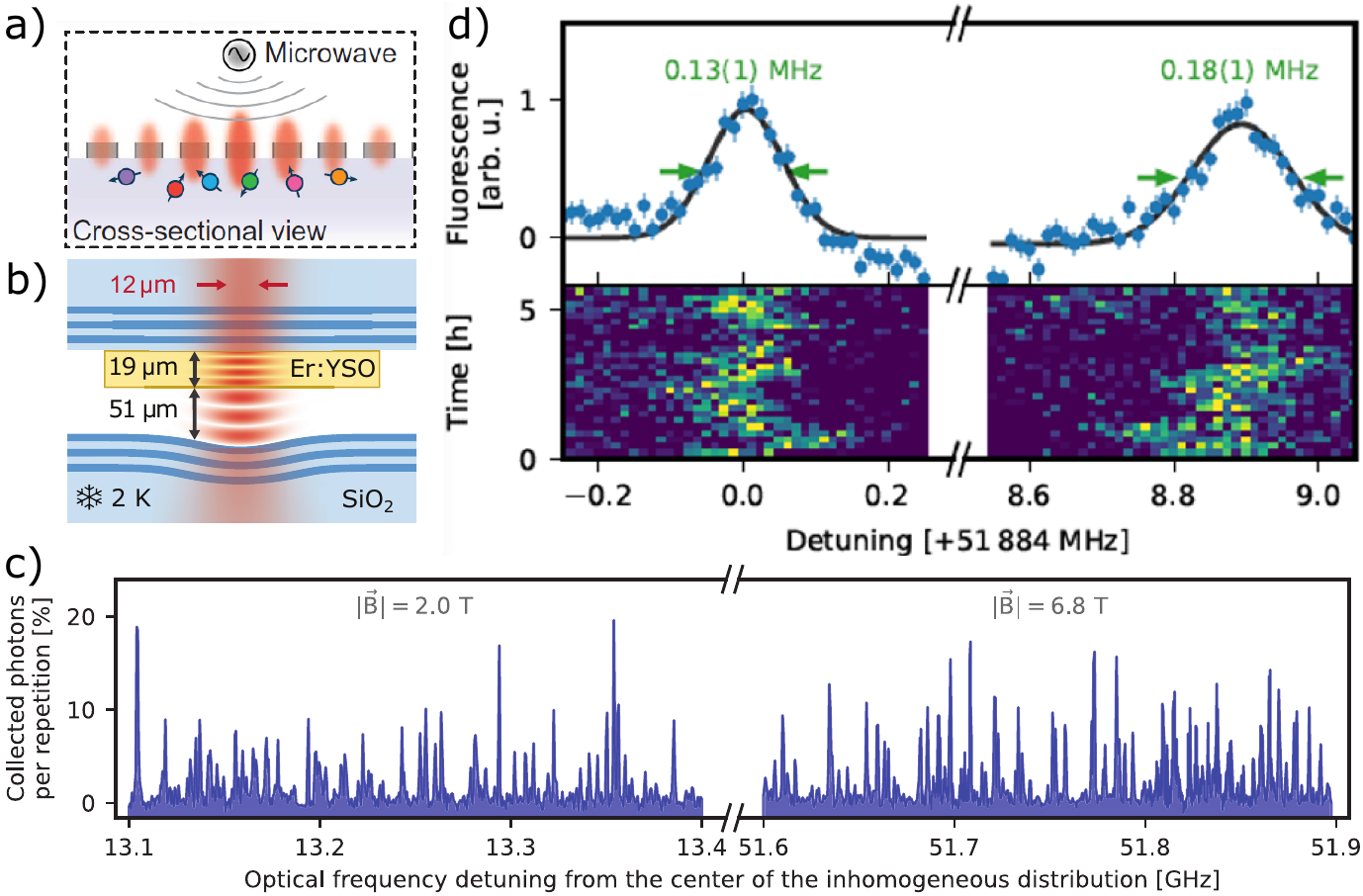}
\caption{   \label{fig_RareEarths}
\textbf{Quantum network node based on rare-earth dopants.}  a) Cross-sectional view of a silicon (grey rectangles) photonic-crystal resonator, fabricated on top of a YSO crystal (purple) by a stamping technique. Individual rare-earth dopants (colored spin symbols) in the evanescent field of the resonator (red ellipsoids) can be addressed individually via their different optical transition frequency. From \cite{chen_parallel_2020}. b) Fabry-Perot resonator. Rare-earth dopants are integrated into a $19\,\si{\micro\meter}$ thin YSO membrane placed between two dielectric mirrors (light and dark blue layers), one of which has a depression to form a stable cavity mode (red ellipsoids). From \cite{ulanowski_spectral_2021}. c) Spectral multiplexing. The resonant emission frequency of the dopants (colored spin symbols in a) depends on the local crystalline environment and can thus differ by several $\si{\giga\hertz}$, with a typical FWHM inhomogeneous linewidth of a few hundred $\si{\mega\hertz}$. Scanning the optical frequency detuning of the excitation laser thus enables the spectral resolution and coherent control of hundreds of individual dopants (fluorscence peaks) in a few-wavelength-scale volume. d) Spectral stability of two resonator-integrated Erbium dopants in YSO, measured at the same time. Only small, uncorrelated fluctuations of the emission maximum (bottom) are observed over several hours. The time-integrated spectral diffusion linewidth (top) of $\sim\SI{0.2}{\mega\hertz}$ is consistent with the expected broadening caused by the bath of Y nuclear spins, and may thus be reduced or eliminated in other hosts materials. From \cite{ulanowski_spectral_2021}.}
\end{figure*}

A common host crystal that fulfills most of these requirements is yttrium orthosilicate (YSO). Other materials can also be favorably used depending on the application. In many hosts, the crystal field splittings are on the order of a few $\si{\tera\hertz}$, such that only the lowest manifold is significantly populated at liquid helium temperature. Upon optical excitation and decay, higher lying crystal field levels can be populated, but they quickly relax to the ground state by phonon emission.

The rare-earth dopants can be further classified into Kramers (typically Ce, Nd, Er, Yb) and non-Kramers (Pr, Eu, Tm) ions with odd and even number of 4f electrons, respectively. The degeneracy of the spin degree of freedom of Kramers ions with their single unpaired electron is lifted in an external magnetic field. The electronic state can then be modeled as a two-level system, i.e. an ideal qubit, in both the ground and optically excited state manifold \cite{thiel_rare-earth-doped_2011}. Because of the large angular momentum of electrons in the 4f shell, the effective g-factor can be very large. This not only makes Kramers ions well-suited for molecular magnets \cite{coronado_molecular_2020}, but could also allow for sensitive magnetic field sensors, microwave quantum memories \cite{probst_microwave_2015} and microwave-to-optical transducers \cite{bartholomew_-chip_2020}. However, the strong and anisotropic interactions between Kramers dopants pose a challenge in this respect, as they can limit the spin lifetime \cite{car_optical_2019} and coherence even when applying tailored dynamical decoupling sequences \cite{merkel_dynamical_2021}.

Thus, using the electronic spin of Kramers dopants in quantum network nodes only seems promising at ultra-low concentrations \cite{cova_farina_coherent_2021, dantec_twenty-threemillisecond_2021}. As an alternative, long-lived quantum states can be encoded in the nuclear rather than electronic spin of the dopant \cite{rancic_coherence_2018, ortu_simultaneous_2018, rakonjac_long_2020, kindem_control_2020, ruskuc_nuclear_2022}. To this end, the electronic spin of Kramers dopants can also be frozen to the ground state at low temperature ($\lesssim 2\,\si{\kelvin}$) and large magnetic fields ($\gtrsim 3\,\si{\tesla}$). In this way, second-long coherence has been obtained \cite{rancic_coherence_2018}, and further improvement is expected in other hosts or at lower temperature.

In non-Kramers systems with their quenched electronic magnetic moment, even longer coherence times have been achieved, with the current record of $6\,\si{\hour}$ for the hyperfine states in Eu:YSO \cite{zhong_optically_2015}. As this host exhibits a large number of nuclear spins, achieving such long coherence relies on two effects: First, the direction and amplitude of an external magnetic field is tuned such that the hyperfine transition frequency is first-order insensitive to magnetic field fluctuations \cite{langer_long-lived_2005}, which is possible even at zero external field with Kramers dopants  \cite{ortu_simultaneous_2018, rakonjac_long_2020, kindem_control_2020}. Second, the dynamics of the nuclear spin bath is slowed down in the "frozen core" that is generated around a rare-earth impurity by its magnetic moment \cite{geschwind_electron_1972}. The detrimental effect of the remaining slow nuclear spin bath dynamics, and other effects such as temperature drifts, can be alleviated by dynamical decoupling \cite{suter_colloquium_2016}. 

Using the above-mentioned techniques, exceptional coherence of both ground-state and optical transitions can be obtained, offering great promise for the implementation of quantum networks. There is only one major challenge in this respect: The protected intra-4f transitions of the rare earths have only weak dipole moments. In free space, they are forbidden by symmetry, and even in crystals the observed lifetimes are typically in the range of milliseconds. For this reason, early quantum network experiments with rare-earth dopants have used large ensembles, as discussed in Sec. \ref{sec_ensembles}. This  has allowed for the implementation of efficient and broadband quantum memories \cite{lvovsky_optical_2009, afzelius_quantum_2015} that can store entangled photons \cite{clausen_quantum_2011, saglamyurek_broadband_2011}  and offer a large multiplexing capacity \cite{tittel_photon-echo_2010, afzelius_quantum_2015} that can be utilized in tailored quantum repeater protocols \cite{sinclair_spectral_2014}.

Still, in spite of low count rates, also single dopants \cite{kolesov_optical_2012, utikal_spectroscopic_2014} and nuclear spins in their proximity \cite{kornher_sensing_2020} have been detected. To use such system for quantum networks, improving the spin readout and photon generation speed is highly desirable. This can be achieved by integrating the emitters into optical resonators. Recent experiments with nanophotonic structures have resolved single dopants \cite{dibos_atomic_2018, zhong_optically_2018, xia_high-speed_2021}, implemented single-shot readout \cite{raha_optical_2020, kindem_control_2020} and nuclear spin registers \cite{ruskuc_nuclear_2022}, and demonstrated frequency-domain multiplexing and simultaneous control of  several dopants \cite{chen_parallel_2020, ulanowski_spectral_2021}. In these experiments, Purcell enhancement factors between 100 and 1000 have been achieved, reducing the optical lifetime to a few $\si{\micro\second}$. This is short compared to the time it takes to transmit photons to remote quantum network nodes, such that it will not limit the achievable rate in remote entanglement experiments. To implement this basic quantum network functionality, the transition frequency of the emitters has to be stable, which is difficult in nanostructures, as the proximity of charge traps at the interface can lead to considerable spectral diffusion linewidths. Using Er:YSO in close proximity to a nanophotonic silicon resonator, $\sim10\,\si{\mega\hertz}$ have been measured \cite{dibos_atomic_2018}. In sites that lack a linear Stark shift \cite{macfarlane_optical_2007}, narrower lines have been observed, e.g. $\sim 1\,\si{\mega\hertz}$ with Yb:YVO, which is close to the lifetime limit in this particular experiment \cite{kindem_control_2020}.

Another approach to obtain large Purcell enhancement is the integration of rare-earth dopants into Fabry-Perot resonators. In contrast to experiments with nanocrystals \cite{casabone_cavity-enhanced_2018, casabone_dynamic_2021}, the use of polished crystalline membranes allows for considerable Purcell enhancement while preserving the optical coherence and spectral stability observed in bulk materials \cite{merkel_coherent_2020}. Recent progress in this setup is shown in Fig. \ref{fig_RareEarths}c-d \cite{ulanowski_spectral_2021}. When operating at a large detuning from the center of the inhomogeneous line, single erbium dopants are spectrally resolved, albeit $\sim10^7$ dopants fall within the cavity mode, and $\sim10^4$ within a diffraction limited volume. The observed Purcell enhancement reaches $\sim 70$-fold, depending on the position of the dopants in the standing-wave cavity mode (panel b). The frequency of the individual peaks is stable over several hours (panel d), with an averaged FWHM of $<0.2\,\si{\mega\hertz}$. These narrow lines allow for resolving and controlling around $10^3$ dopants when fast resonator tuning \cite{casabone_dynamic_2021} is implemented.

The remaining broadening is explained by the coupling of the electronic spin to the nuclear spin bath \cite{merkel_enhancing_2021}. Thus, a considerable improvement is expected when using the isotope $^{167}\text{Er}$ at a magnetic field insensitive point \cite{ortu_simultaneous_2018, rakonjac_long_2020}. Alternatively, different host materials that have only a small abundance of nuclear magnetic moments can be considered. Recently studied materials include $\text{TiO}_2$ \cite{phenicie_narrow_2019}, calcium tungstate \cite{dantec_twenty-threemillisecond_2021}, and crystalline silicon \cite{yin_optical_2013, weiss_erbium_2021, berkman_sub-megahertz_2021, gritsch_narrow_2021}. The latter seems particularly promising, as isotopically purified material can be epitaxially grown by chemical vapor deposition (CVD) on a wafer scale \cite{mazzocchi_99992_2019}.

Recent experiments in CVD silicon with natural isotope abundance have revealed narrow inhomogeneous ($< \SI{1}{\giga\hertz}$) and homogeneous ($\lesssim 20\,\si{\kilo\hertz}$) linewidths of erbium dopants in particular lattice sites \cite{gritsch_narrow_2021} - at a par with established host materials such as YSO. Remarkably, because of its high refractive index \cite{de_vries_resonant_1998}, the radiative lifetime in silicon can be almost 50 times shorter than in YSO, enhancing the expected rate in quantum network experiments. When integrating isotopically purified membranes into Fabry-Perot resonators, a large number of dopants with negligible spectral diffusion may be controlled. Combined with its emission at telecommunication frequency and the prospect for second-long ground-state coherence \cite{rancic_coherence_2018}, this makes such systems a promising platform for the implementation of global quantum networks and quantum repeaters.

\section{Summary and outlook}

The integration of single emitters into low-loss optical resonators has unique potential for the realization of scalable quantum networks. First steps into this direction have been taken in several experimental platforms, which have demonstrated the successful initialization, control, readout and remote entanglement of spin qubits based on efficient spin-photon interfaces. Still, scaling the demonstrated elementary quantum links to a network with many nodes that are distributed over global distances poses a formidable challenge. While many concepts that allow for such networks have been developed \cite{muralidharan_optimal_2016, wehner_quantum_2018}, the experimental requirements of high efficiency and almost $100\,\%$ fidelity  are difficult to achieve in all investigated physical platforms, and will therefore require a considerable engineering effort. Still, even with present experimental imperfections, the realization of a prototype quantum repeater seems within reach \cite{rozpedek_parameter_2018}.

Using such systems outside of the lab, e.g. in a global communication scenario, will only be possible if the devices are robust and cost-effective. Thus, the integration of the presented quantum network nodes with on-chip photonics \cite{wang_integrated_2020} will likely receive growing attention. Then, based on the current optical fiber infrastructure, the maximum separation of quantum repeater nodes will have to be on the order of $100\,\si{\kilo\meter}$, a distance after which $99\,\%$ of the photons have been lost. Covering continental distances with an equally spaced network of high-vacuum chambers or closed-cycle $^4\text{He}$ cryostats at such spacing seems feasible. This would be sufficient to provide users that have limited quantum processing capacity, e.g. only photodetectors and phase or polarization modulators, access to quantum network resources. Bridging the gaps between continents, however, seems more difficult and may favor the use of quantum satellites \cite{yin_satellite-based_2017} or drones \cite{liu_drone-based_2020} rather than fiber-based links.

Going beyond point-to-point connections to generate entangled states of many nodes will require the implementation of entanglement distillation or quantum error correction. In the latter, current topological codes allow for error probabilities of a few per cent \cite{fowler_surface_2012, nickerson_topological_2013}, but then require an impractical overhead. Therefore, increasing the fidelity of remote entanglement far beyond what is achieved in current experiments will be paramount. To this end, the spectral stability of the emitters and resonators will have to be improved, in particular for solid-state qubits. Alternatively, much larger Purcell enhancement may be targeted. 

If successful, the implementation of large quantum networks will open the door for novel fundamental tests \cite{brunner_bell_2014, pikovski_universal_2015} at the forefront of contemporary quantum science. In addition, they will enable numerous applications. In particular, the upscaling of quantum computers may be based on a modular architecture \cite{monroe_scaling_2013, awschalom_quantum_2013, kinos_roadmap_2021}, similar to the distribution of information processing among different components in classical high-performance computers and data centers. To this end, boosting the rate of remote entanglement to approach that of local quantum gates is highly desirable. This will further stimulate the research into optimized materials systems and resonator-integrated qubit platforms in the coming decades.

\section{Acknowledgements}
I acknowledge discussions with Benjamin Merkel and Andreas Gritsch, and funding from the European Research Council (ERC) under the European Union's Horizon 2020 research and innovation programme (grant agreement No 757772), and from the Deutsche Forschungsgemeinschaft (DFG, German Research Foundation) under Germany's Excellence Strategy - EXC-2111 - 390814868.

\bibliography{bibliography.bib}

\end{document}